\documentclass[letterpaper,12pt,notitlepage]{article}
\usepackage{amsmath,amssymb,rotating,amsthm,delarray,dcolumn,graphics,epsfig,soul,amsthm,longtable, lscape, multirow,array, caption,anyfontsize, float}
\usepackage[usenames,dvipsnames]{color}
\usepackage{booktabs,siunitx}
\usepackage{verbatim}
\usepackage{setspace}
\usepackage{footmisc}
\usepackage{rotating}
\usepackage{setspace}
\usepackage{booktabs,caption}
\usepackage[flushleft]{threeparttable}
\usepackage{arydshln}
\usepackage[titletoc,toc,page]{appendix}
\usepackage{color}
\renewcommand{\thetable}{\Roman{table}}
\usepackage{enumitem,kantlipsum}

\usepackage{graphicx}
\newsavebox\CBox

\makeatletter
\def\@makefnmark{%
  \leavevmode
  \raise.9ex\hbox{\fontsize\sf@size\z@\normalfont\tiny\@thefnmark}}
\makeatother

\theoremstyle{definition}

\usepackage{lipsum}
\usepackage[bottom=-0.5in,top=1.0in, left = 1in, right = 1in]{geometry}
\usepackage[english]{babel}
\usepackage{authblk,indentfirst}
\makeatletter
\renewcommand\@biblabel[1]{}
\makeatother
\renewcommand
\baselinestretch{1} 

\usepackage{sectsty}
\sectionfont{\centering}
\subsectionfont{\normalfont\itshape}
\subsubsectionfont{\normalfont\itshape\centering}

\addto{\captionsenglish}{}

\begin{document}

\begin{titlepage}
\title{\bf False (and Missed) Discoveries in Financial Economics  }

\date{}

\author{
\small {\bf CAMPBELL R. HARVEY and YAN LIU}\thanks{
Harvey is with Duke University and National Bureau of Economic Research. Liu is with Purdue University. Send correspondence to: Yan Liu, Krannert School of Management, Purdue University, West Lafayette, IN 47905. Phone: +1 919.428.1118, E-mail: liu2746@purdue.edu. We appreciate the comments of Stefan Nagel, two anonymous referees, as well as Laurent Barras, Claude Erb, Juhani Linnainmaa, Michael Weber, and seminar participants at Rice University, University of Southern California, University of Michigan, University of California at Irvine, Hanken School of Economics, Man-Numeric, Research Affiliates, and the 2018 Western Finance Association meetings in San Diego. Kay Jaitly provided editorial assistance.}} \vspace{-7.5pt}
\vspace{4mm}
\maketitle
\vspace{-1cm}
\begin{abstract}
\small
\singlespacing
\noindent Multiple testing plagues many important questions in finance such as fund and factor selection. We propose a new way to calibrate both Type I and Type II errors. Next, using a double-bootstrap method, we establish a $t$-statistic hurdle that is associated with a specific false discovery rate (e.g., 5\%). We also establish a hurdle that is associated with a certain acceptable ratio of misses to false discoveries (Type II error scaled by Type I error), which effectively allows for differential costs of the two types of mistakes. Evaluating current methods, we find that they lack power to detect outperforming managers. 
\end{abstract}
\vspace{0cm}

\renewcommand{\baselinestretch}{1.0}
\normalsize \maketitle \thispagestyle{empty}
\end{titlepage}

\setlength{\textheight}{8.95in}
\renewcommand{\baselinestretch}{1.5}
\setcounter{page}{1}

\onehalfspacing
%
\noindent In manager selection (or, equivalently, the selection of factors or trading strategies), investors can make two types of mistakes. The first involves selecting a manager who turns out to be unskilled --- this is a Type I error, or a false positive.\footnote{Throughout our paper, we follow the empirical literature on performance evaluation and associate manager skill with alpha. In particular, we take skilled managers to be those that generate positive alpha. Our notion of manager skill is thus different from that in Berk and Green (2004), where skilled managers generate a zero net alpha in equilibrium.} The second error is not selecting or missing a manager that the investor thought was unskilled but was not --- this is a Type II error, or a false negative. Both types of errors display economically important variation. On the false positives side, for instance, one manager might slightly underperform while another manager might have a large negative return. Moreover,
the cost of a Type II error is likely different from the cost of a Type I error, with the costs depending on the specific decision at hand. However, while an investor may want to pick managers using the criterion that Type I errors are, say, five times more costly than Type II errors, current tools do not allow for such a selection criterion. On the one hand, current methods ignore the Type II error rate, which may lead us to miss outperforming managers. On the other hand, it is difficult to simply characterize Type I errors because of multiple testing --- using a single-hypothesis testing criterion (e.g., two standard errors from zero) will lead to massive Type I errors because, when there are thousands of managers, many will look good (i.e., appear to outperform) purely by luck. Statisticians have suggested a number of fixes that take multiple testing into account. For example, the simplest is the Bonferroni correction, which multiplies each manager's $p$-value by the number of managers. But this type of correction does not take the covariance structure into account. Further, it is not obvious what the Type I error rate would be after implementing the correction. We know that the error rate would be less than that under a single testing criterion –- but how much less?

 
In this paper we propose a different approach. Using actual manager data, we first determine the performance threshold that delivers a particular Type I error rate (e.g., 5\%). We next characterize the Type II error rate associated with our optimized Type I error rate. It is then straightforward to scale the Type II error by the Type I error and solve for the cutoff that produces the desired trade-off of false negatives and false positives.

Our focus on both Type I and Type II errors echoes recent studies in economics that highlight the importance of examining test power. For example, in a survey of a large number of studies, Ioannidis, Stanley, and Doucouliagos (2017) show that 90\% of results in many research areas are under powered, leading to an exaggeration of the results. Ziliak and McCloskey (2004) further show that only 8\% of the papers published in the \emph{American Economic Review} in the 1990s consider test power. The question of test power thus represents one more challenge to research practices common in economics research (Leamer (1983), De Long and Lang (1992), Ioannidis and Doucouliagos (2013), Harvey and Liu (2013), Harvey, Liu, and Zhu (2016), Harvey (2017)).

Why is test power important for research in financial economics? On the one hand, when a study's main finding is the non existence of an effect (i.e., the null hypothesis is not rejected), test power directly affects the credibility of the finding because it determines the probability of not rejecting the null hypothesis when the effect is true. For example, in one of our applications, we show that existing studies lack power to detect outperforming mutual funds. On the other hand, when the main finding is the rejection of the null hypothesis (i.e., the main hypothesis), this finding often has to survive against alternative hypotheses (i.e., alternative explanations for the main finding). Low test power for alternative explanations generates a high Type I error rate for the main hypothesis (Ioannidis (2005)).

%

Our paper addresses the question of test power in the context of multiple tests. Our contribution is threefold. First, we introduce a framework that offers an intuitive definition of test power. Second, we employ a double-bootstrap approach that can flexibly (i.e., specific to a particular data set) estimate test power. Finally, we illustrate how taking test power into account can materially change our interpretation of important research findings in the current literature.

In a single-hypothesis test, the Type II error rate at a particular parameter value (in our context, the performance metric for the manager) is calculated as the probability of failing to reject the null hypothesis at this value. In multiple tests, the calculation of the Type II error rate is less straightforward because, instead of a single parameter value, we need to specify a vector of non zero parameters, where each parameter corresponds to a single test under the alternative hypothesis.

We propose a simple strategy to estimate the Type II error rate. Assuming that a fraction $p_0$ of managers have skill, we adjust the data so that $p_0$ of managers have skill (with their skill level set at the in-sample estimate) and the remaining $1-p_0$ of managers have no skill (with their skill level set to a zero excess return or alpha). By bootstrapping from these adjusted data, we evaluate the Type II error rate through simulations. Our method thus circumvents the difficulty of specifying the high-dimensional parameter vector under the alternative hypothesis. We set the parameter vector at what we consider a reasonable value --- the in-sample estimate corresponding to a certain $p_0$. In essence, we treat $p_0$ as a sufficient statistic, which helps estimate the Type II error rate. We interpret $p_0$ from both a frequentist and a Bayesian perspective.

Our strategy is related to the bootstrap approach in performance evaluation proposed by Kosowski et al. (2006, KTWW) and Fama and French (2010).\footnote{See Harvey and Liu (2019) for another application of the bootstrap approach to the test of factor models.} These papers use a single-bootstrap approach to adjust for multiple testing. In particular, under the assumption of no skill for all funds ($p_0=0$), they demean the data to create a ``pseudo" sample, $Y_0$, for which $p_0 = 0$ holds true in-sample. They then bootstrap $Y_0$ to test the overall hypothesis that all funds have zero alpha. Because we are interested in both the Type I and the Type II error rates associated with a given testing procedure (including those of KTWW and Fama and French (2010)), our method uses two rounds of bootstrapping. For example, to estimate the Type I error rate of Fama and French (2010), we first bootstrap $Y_0$ to create a perturbation, $Y_i$, for which the null hypothesis is true. We then apply Fama and French (2010) (i.e., second bootstrap) to each $Y_i$ and record the testing outcome ($h_i=1$ if rejection). We estimate the Type I error rate as the average $h_i$. The Type II error rate can be estimated in similar fashion.


After introducing our framework, we turn to two empirical applications to illustrate how our framework helps address important issues related to Type I and Type II errors associated with multiple tests. We first apply our method to the selection of two sets of investment factors. The first set includes hundreds of backtested factor returns. For a given $p_0$, our method allows investors to measure the Type I and Type II error rates for these factors and thus make choices that strike a balance between Type I and Type II errors. When $p_0$ is uncertain, investors can use our method to evaluate the performance of existing multiple-testing adjustments and select the adjustment that works well regardless of the value of $p_0$ or for a range of $p_0$ values. Indeed, our application shows that multiple-testing methods usually follow an ordering (best to worst) in performance, regardless of the value of $p_0$.

The second set of investment factors includes around 18,000 anomaly strategies constructed and studied by Yan and Zheng (2017). Relying on the Fama and French (2010) approach to adjust for multiple testing, Yan and Zheng (2017) claim that a large fraction of the 18,000 anomalies in their data are true and conclude that there is widespread mispricing. We use our model to estimate the error rates of their approach and obtain results that are inconsistent with their narrative.

We next apply our approach to performance evaluation, revisiting the problem of determining whether mutual fund managers have skill. In particular, we use our double-bootstrap technique to estimate the Type I and Type II error rates of the popular Fama and French (2010) approach. We find that their approach lacks power to detect outperforming funds. Even when a significant fraction of funds are outperforming, and the returns that these funds generate in the actual sample are large, the Fama and French (2010) method may still declare, with a high probability, a zero alpha across all funds. Our result thus calls into question their conclusions regarding mutual fund performance and helps reconcile the difference between KTWW and Fama and French (2010).

Our paper is not alone in raising the question of power in performance evaluation. Ferson and Chen (2017), Andrikogiannopoulou and Papakonstantinou (2019), and Barras, Scaillet, and Wermers (2018) focus on the power of applying the false discovery rate approach of Barras, Scaillet, and Wermers (2010) in estimating the fraction of outperforming funds. Our paper differs by proposing a non parametric bootstrap-based approach to systematically evaluate test power. We also apply our method to a wide range of questions in financial economics, including the selection of investment strategies, identification of equity market anomalies, and evaluation of mutual fund managers. Chordia, Goyal, and Saretto (2020) study an even larger collection of anomalies than Yan and Zheng (2017) and use several existing multiple-testing adjustment methods to estimate the fraction of true anomalies. In contrast to Chordia, Goyal, and Saretto (2020), we compare different methods using our bootstrap-based approach. In particular, we show exactly what went wrong with the inference in Yan and Zheng (2017).

Our paper also contributes to the growing literature in finance that applies multiple-testing techniques to related areas in financial economics (see, for example, Harvey, Liu, and Zhu (2016), Harvey (2017), Chordia, Goyal, and Saretto (2020), Barras (2019), Giglio, Liao, and Xiu (2018)). One obstacle for this literature is that, despite the large number of available methods developed by the statistics literature, it is unclear which method is most suitable for a given data set. We provide a systematic approach that offers data-driven guidance on the relative performance of multiple-testing adjustment methods.

Our paper is organized as follows. In Section I, we present our method. In Section II, we apply our method to the selection of investment strategies and mutual fund performance evaluation. In Section III, we offer some concluding remarks.

\section{Method}

\subsection{Motivation: A Single Hypothesis Test}
\indent Suppose we have a single hypothesis to test and the test corresponds to the mean of a univariate variable $X$. For a given testing procedure (e.g., the sample mean $t$ test), at least two metrics are important for gauging the performance of the procedure. The first is the Type I error rate, which is the probability of incorrectly rejecting the null when it is true (a false positive), and the other is the Type II error rate, which is probability of incorrectly declaring insignificance when the alternative hypothesis is true (a false negative). The Type II error rate is also linked to test power (i.e., power $= 1-$ Type II error rate), which is the probability of correctly rejecting the null when the alternative hypothesis is true.

In a typical single-hypothesis test, we try to control the Type I error rate at a certain level (e.g., 5\% significance) while seeking methods that generate a low Type II error rate or, equivalently, high test power. To evaluate the Type I error rate, we assume that the null is true (i.e., $\mu_0 = 0$) and calculate the probability of a false discovery. For the Type II error rate, we assume a certain level (e.g., $\mu_0 = 5\%$) for the parameter of interest (i.e., the mean of $X$) and calculate the probability of a false negative as a function of $\mu_0$.

In the context of multiple hypothesis testing, the evaluation of Type I and Type II error rates is less straightforward for several reasons. First, for the definition of the Type I error rate, the overall hypothesis that the null hypothesis holds for each individual test may be too strong to be realistic for certain applications. As a result, we often need alternative definitions of Type I error rates that apply even when some of the null hypotheses may not be true.\footnote{One example is the False Discovery Rate, as we shall see below. } For the Type II error rate, its calculation generally depends on the parameters of interest, which is a high-dimensional vector as we have multiple tests. As a result, it is not clear what value for this high-dimensional vector is most relevant for the calculation of the Type II error rate.

Given the difficulty in determining the Type II error rate, current multiple-testing adjustments often focus only on Type I errors. For example, Fama and French (2010) look at mutual fund performance and test the overall null hypothesis of a zero alpha for all funds. They do not assess the performance of their method when the alternative is true, that is, the probability of incorrectly declaring insignificant alphas across all funds when some funds display skill. As another example, the multiple-testing adjustments studied in Harvey, Liu, and Zhu (2016) focus on the Type I error defined by either the family-wise error rate (FWER), which is the probability of making at least one false discovery, or the false discovery rate (FDR), which is the expected fraction of false discoveries among all discoveries. Whether these methods have good performance in terms of Type I error rates and what the implied Type II error rates would be thus remain open questions.

Second, while we can often calculate Type I and Type II error rates analytically under certain assumptions for a single hypothesis test, such assumptions become increasingly untenable when we have many tests. For example, it is difficult to model cross-sectional dependence when we have a large collection of tests.


Third, when there are multiple tests, even the definitions of Type I and Type II error rates become less straightforward. While traditional multiple-testing techniques apply to certain definitions of Type I error rates such as FWER or FDR, we are interested in a general approach that allows us to evaluate different measures of the severity of false positives and false negatives. For example, while the FWER from the statistics literature has been applied by Harvey, Liu, and Zhu (2016) to evaluate strategies based on anomalies, an odds ratio that weighs the number of false discoveries against the number of misses may be more informative for the selection of investment strategies as it may be more consistent with the manager's objective function.

Motivated by these concerns, we develop a general framework that allows us to evaluate error rates in the presence of multiple tests. First, we propose a simple metric to summarize the information contained in the parameters of interest and to evaluate Type I and Type II error rates. In essence, this metric reduces the dimensionality of the parameters of interest and allows us to evaluate error rates around what we consider a reasonable set of parameter values. Second, we evaluate error rates using a bootstrap method, which allows us to capture cross-sectional dependence nonparametrically. Because our method is quite flexible in terms of how we define the severity of false positives and false negatives, we are able to evaluate error rate definitions that are appropriate for a diverse set of finance applications.

\subsection{Bootstrapped Error Rates under Multiple Tests}
\indent To ease our exposition, we describe our method in the context of testing the performance of many trading strategies. Suppose we have $N$ strategies and $D$ time periods. We arrange the data into a $D \times N$ data matrix $X_0$.

Suppose one believes that a fraction $p_0$ of the $N$ strategies are true. We develop a simulation-based framework to evaluate error rates related to multiple hypothesis testing for a given $p_0$.

There are several ways to interpret $p_0$. When $p_0=0$, no strategy is believed to be true, which is the overall null hypothesis of a zero return across all strategies. This hypothesis can be economically important. For example, KTWW and Fama and French (2010) examine this hypothesis for mutual funds to test market efficiency. We discuss this hypothesis in detail in Section I.D when we apply our method to Fama and French (2010).

When $p_0>0$, some strategies are believed to be true. In this case, $p_0$ can be thought of as a plug-in parameter --- similar to the role of $\mu_0$ in a single test as discussed in Section I.A  --- that helps us measure the error rates in the presence of multiple tests. As we discuss above in the context of multiple tests, in general one needs to make assumptions about the values of the population statistics (e.g., the mean return of a strategy) for all of the strategies believed to be true in order to determine the error rates. However, in our framework we argue that $p_0$ serves as a single summary statistic that allows us to effectively evaluate error rates without having to condition on the values of the population statistics. As a result, we offer a simple way to extend the error rate analysis for a single hypothesis test to the concept of multiple hypothesis tests.

Note that by choosing a certain $p_0$, investors are implicitly taking a stand on the plausible economic magnitudes of alternative hypotheses. For example, investors may believe that strategies with a mean return above 5\% are likely to be true, resulting in a corresponding $p_0$. While this is one way to rationalize the choice of $p_0$ in our framework, our model allows us to evaluate the implications of not only $p_0$ or the 5\% cutoff but the entire distribution of alternatives on error rates.

The parameter $p_0$ also has a Bayesian interpretation. Harvey, Liu, and Zhu (2016) present a stylized Bayesian framework for multiple hypothesis testing in which multiple-testing adjustment is achieved indirectly through the likelihood function. Harvey (2017) instead recommends the use of the minimum Bayes factor, which builds on Bayesian hypothesis testing but abstracts from the prior specification by focusing on the prior that generates the minimum Bayes factor. Our treatment of $p_0$ lies between Harvey, Liu, and Zhu (2016) and Harvey (2017) in the sense that while we do not go as far as making assumptions on the prior distribution of $p_0$ that is later fed into the full-blown Bayesian framework as in Harvey, Liu, and Zhu (2016), we deviate from the Harvey (2017) assumption of a degenerate prior (i.e., the point mass that concentrates on the parameter value that generates the minimum Bayes factor) by exploring how error rates respond to changes in $p_0$. While we do not attempt to pursue a complete Bayesian solution to multiple hypothesis testing,\footnote{Storey (2003) provides a Bayesian interpretation of the positive FDR. Scott and Berger (2006) include a general discussion of Bayesian multiple testing. Harvey, Liu, and Zhu (2016) discuss some of the challenges in addressing multiple testing within a Bayesian framework. Harvey et al. (2019) present a full-blown Bayesian framework to test market efficiency. } the sensitivity of error rates to changes in $p_0$ that we highlight in this paper are important ingredients to both Bayesian and frequentist hypothesis testing.

Although the choice of $p_0$ is inherently subjective, we offer several guidelines as to the selection of an appropriate $p_0$. First, examination of the summary statistics for the data can help narrow the range of reasonable priors.\footnote{See Harvey (2017) for examples of Bayes factors that are related to data-driven priors.} For example, about 22\% of strategies in our CAPIQ sample (see Section II.A for details on these data) have a $t$-statistic above 2.0. This suggests that, at the 5\% significance level, $p_0$ is likely lower than 22\% given the need for a multiple-testing adjustment. Second, it may be a good idea to apply prior knowledge to elicit $p_0$. For example, researchers with a focus on quantitative asset management may have an estimate of the success rate of finding a profitable investment strategy that is based on past experience. Such an estimate can guide their choices of $p_0$. Finally, while in principle researchers can pick any $p_0$ in our model, in Section II.B we present a simulation framework that helps gauge the potential loss from applying different priors.

For a given $p_0$, our method starts by choosing $p_0\times N$ strategies that are deemed to be true. A simple way to choose these strategies is to first rank the strategies by their $t$-statistics and then choose the top $p_0\times N$ with the highest $t$-statistics. While this approach is consistent with the idea that strategies with higher $t$-statistics are more likely to be true, it ignores the sampling uncertainty in ranking the strategies. To take this certainty into account, we perturb the data and rank the strategies based on the perturbed data. In particular, we bootstrap the time periods and create an alternative panel of returns, $X_i$ (note that the original data matrix is $X_0$). For $X_i$, we rank its strategies based on their $t$-statistics. For the top $p_0\times N$ strategies with the highest $t$-statistics, we find the corresponding strategies in $X_0$. We adjust these strategies so that their in-sample means are the same as the means for the top $p_0\times N$ strategies in $X_i$.\footnote{Alternatively, if a factor model is used for risk adjustment, we could adjust the intercepts of these strategies after estimating a factor model so that the adjusted intercepts are the same as those for the top $p_0\times N$ strategies in $X_i$.} We denote the data matrix for these adjusted strategies by $X^{(i)}_{0,1}$. For the remaining strategies in $X_0$, we adjust them so they have a zero in-sample mean (denote the data matrix for these adjusted strategies by $X^{(i)}_{0,0}$). Finally, we arrange $X^{(i)}_{0,1}$ and $X^{(i)}_{0,0}$ into a new data matrix $Y_i$ by concatenating the two data matrices. The data in $Y_i$ will be the hypothetical data that we use to perform our follow-up error rate analysis, for which we know exactly which strategies are true and which strategies are false.


Our strategy of constructing a ``pseudo" sample under the alternative hypothesis (i.e., some strategies are true) is motivated by the bootstrap approach proposed by the mutual fund literature. In particular, KTWW perform a bootstrap analysis at the individual fund level to select ``star" funds. Fama and French (2010) look at the cross-sectional distribution of fund performance to control for multiple testing. Both papers rely on the idea of constructing a ``pseudo" sample of fund returns for which the null hypothesis of zero performance is known to be true. We follow their strategies by constructing a similar sample for which some of the alternative hypotheses are known to be true, with their corresponding parameter values (i.e., strategy means) set at plausible values, in particular, their in-sample means associated with our first-stage bootstrap.

Notice that due to sampling uncertainty, what constitutes alternative hypotheses in our first-stage bootstrap may not correspond to the true alternative hypotheses for the underlying data-generating process. In particular, strategies with a true mean return of zero in population may generate an inflated mean after the first-stage bootstrap and thus be falsely classified as alternative hypotheses. While existing literature offers several methods to shrink the in-sample means (e.g., Jones and Shanken (2005), Andrikogiannopoulou and Papakonstantinou (2016), and Harvey and Liu (2018)), we do not employ them into our current paper. In a simulation study in which we evaluate the overall performance of our approach, we treat the misclassification of hypotheses as one potential factor that affects our model performance. Under realistic assumptions for the data-generating process, we show that our method performs well despite the potential misclassification of alternative hypotheses.

For $Y_i$, we bootstrap the time periods $J$ times to evaluate the error rates for a statistical procedure, such as a fixed $t$-statistic threshold (e.g., a conventional $t$-statistic threshold of 2.0) or the range of multiple-testing approaches detailed in Harvey, Liu, and Zhu (2016). By construction, we know which strategies in $Y_i$ are believed to be true (and false), which allows us to summarize the testing outcomes for the $j$-th bootstrap iteration with the vector $\bar{O}^{i,j}= (TN^{i,j}, FP^{i,j}, FN^{i,j}, TP^{i,j})'$, where $TN^{i,j}$ is the number of tests that correctly identify a false strategy as false (true negative), $FP^{i,j}$ is the number of tests that incorrectly identify a false strategy as true (false positive), $FN^{i,j}$ is the number of tests that incorrectly identify a true strategy as false (false negative), and $TP^{i,j}$ is the number of tests that correctly identify a true strategy as true (true positive). Notice that for brevity we suppress the dependence of $\bar{O}^{i,j}$ on the significance threshold (i.e., either a fixed $t$-statistic threshold or the threshold generated by a data-dependent testing procedure). Table \ref{table:classify} illustrates these four summary statistics using a contingency table.

\begin{table}[!h]
\centering 
  \footnotesize\addtolength{\tabcolsep}{-4pt}
  \captionsetup{justification=centering}
  \caption{\textbf{Classifying Testing Outcomes}  }
\captionsetup{width=16.0cm}
\captionsetup{justification=centering}
\caption*{\textmd{\small{A Contingency table for testing outcomes.}} ]}
\small{
\begin{tabular}{cc c cc} 
\hline\hline
Decision &&  Null  && Alternative \\
         && ($H_0$)  && ($H_1$) \\
\hline
Reject  &&  False positive    & & True positive \\
        & & (Type I error)      &&  ($TP^{i,j}$) \\
        & & ($FP^{i,j}$)   &&   \\
\multicolumn{5}{c}{} \\ [-3mm]
Accept  & & True negative     && False negative \\
        & &  ($TN^{i,j}$)                    && (Type II error) \\
        & &     && ($FN^{i,j}$) \\
\hline
  \end{tabular}\par}
\label{table:classify} 
\end{table}

With these summary statistics, we can construct several error rates of interest. We focus on three types of error rates in our paper. The first --- the \textit{realized} false discovery rate (FDR) --- is motivated by the FDR (see Benjamini and Hochberg (1995), Benjamini and Yekutieli (2001), Barras, Scaillet, and Wermers (2010), and Harvey, Liu, and Zhu (2016)) and is defined as
\begin{eqnarray*}
RFDR^{i,j} &=& \left\{
  \begin{array}{rcr}
    & \frac{FP^{i,j}}{FP^{i,j} + TP^{i,j}}  , \ \text{if} \ \ FP^{i,j} + TP^{i,j} >0,  \\
    &  \ \ 0, \ \ \ \ \ \ \text{if} \ \  FP^{i,j} + TP^{i,j} =0, \\
  \end{array}
\right.
\end{eqnarray*}
that is, the fraction of false discoveries (i.e., $FP^{i,j}$) among all discoveries (i.e., $FP^{i,j} + TP^{i,j}$). The expected value of $RFDR$ extends the Type I error rate in a single hypothesis test to multiple tests.

The second type of error rate --- the realized rate of misses (RMISS), sometimes referred to as the false omission rate or false non discovery rate --- is also motivated by the FDR and is defined as
\begin{eqnarray*}
RMISS^{i,j} &=& \left\{
  \begin{array}{rcr}
    & \frac{FN^{i,j}}{FN^{i,j} + TN^{i,j}}  , \ \text{if} \ \ FN^{i,j} + TN^{i,j} >0,  \\
    & \ \ 0, \ \ \ \ \ \ \text{if} \ \  FN^{i,j} + TN^{i,j} =0, \\
  \end{array}
\right.
\end{eqnarray*}
that is, the fraction of misses (i.e., $FN^{i,j}$) among all tests that are declared insignificant (i.e., $FN^{i,j} + TN^{i,j}$). The expected value of $RMISS$ extends the Type II error rate in a single hypothesis test to multiple tests.\footnote{Alternative error rate definitions include \emph{precision} (the ratio of the number of correct positives to the number of all predicted positives, that is, $TP^{i,j}/(FP^{i,j} + TP^{i,j})$) and \emph{recall} (also known as the hit rate or true positive rate; the ratio of the number of true positives to the number of strategies that should be identified as positive, that is, $TP^{i,j}/(TP^{i,j} + FN^{i,j})$). One can also define the FDR as the expected fraction of false discoveries among all tests for which the null is true, which corresponds more closely to the Type I error rate definition for a single test.}

Finally, similar to the concept of the odds ratio in Bayesian analysis, we define the realized ratio of false discoveries to misses (RRATIO) as
\begin{eqnarray*}
RRATIO^{i,j} &=& \left\{
  \begin{array}{rcr}
    & \frac{FP^{i,j}}{FN^{i,j}} , \ \text{if} \ \ FN^{i,j} >0,  \\
    & 0, \ \ \ \  \text{if} \ \  FN^{i,j} =0, \\
  \end{array}
\right.
\end{eqnarray*}
that is, the ratio of false discoveries (i.e., $FP^{i,j}$) to misses (i.e., $FN^{i,j}$).\footnote{In most of our applications there are misses, so the difference between our current definition (i.e., $RRATIO^{i,j} = 0 \ \ \text{if} \ \ FN^{i,j} =0$) and alternative definitions, such as excluding simulation runs for which $FN^{i,j} =0$, is small.}

Notice that by using summary statistics that count the number of occurrences for different types of testing outcomes, we are restricting attention to error rate definitions that depend only on the number of occurrences. Alternative definitions of error rates that involve the magnitudes of the effects being tested (e.g., an error rate that puts a higher weight on a missed strategy with a higher Sharpe ratio) can also be accommodated in our framework.\footnote{See DeGroot (1975), DeGroot and Schervish (2011, chapter 9), and Beneish (1997, 1999). }

Finally, we account for the sampling uncertainty in ranking the strategies and the uncertainty in generating the realized error rates for each ranking by averaging across both $i$ and $j$. Suppose we perturb the data $I$ times, and each time we generate $J$ bootstrapped random samples. We then have
\begin{eqnarray*}
TYPE1 &=& \frac{1}{IJ}\sum_{i=1}^I \sum_{j=1}^J RFDR^{i,j}, \\
TYPE2 &=& \frac{1}{IJ}\sum_{i=1}^I \sum_{j=1}^J RMISS^{i,j}, \\
ORATIO &=& \frac{1}{IJ}\sum_{i=1}^I \sum_{j=1}^J RRATIO^{i,j}. \\
\end{eqnarray*}
We refer to $TYPE1$ as the Type I error rate, $TYPE2$ as the Type II error rate, and $ORATIO$ as the odds ratio (between false discoveries and misses). Notice that similar to $\bar{O}^{i,j}$, our estimated $TYPE1$, $TYPE2$, and $ORATIO$ implicitly depend on the significance threshold.

There are several advantages of $ORATIO$ compared to $TYPE1$ and $TYPE2$. First, $ORATIO$ links Type I and Type II errors by quantifying the chance of a false discovery per miss. For example, if an investor believes that the cost of a Type I error is 10 times that of a Type II error, then the optimal $ORATIO$ should be $1/10$. Second, $ORATIO$ takes the magnitude of $p_0$ into account. When $p_0$ is very small, $TYPE2$ is usually much smaller than $TYPE1$. However, this mainly reflects the rare occurrence of the alternative hypothesis and does not necessarily imply good model performance in controlling $TYPE2$. In this case, $ORATIO$ may be a more informed metric in balancing Type I and Type II errors. While we do not attempt to specify the relative weight between $TYPE1$ and $TYPE2$ (which likely requires specifying a loss function that weighs Type I errors against Type II errors), we use $ORATIO$ as a heuristic to weigh Type I errors against Type II errors.


To summarize, we follow the steps below to evaluate the error rates.

\begin{enumerate}[align=left]
\item[Step I.] Bootstrap the time periods and let the bootstrapped panel of returns be $X_i$. For $X_i$, obtain the corresponding $1\times N$ vector of $t$-statistics $t_i$.
\item[Step II.] Rank the components in $t_i$. For the top $p_0$ of strategies in $t_i$, find the corresponding strategies in the original data $X_0$. Adjust these strategies so they have the same means as those for the top $p_0$ of strategies ranked by $t_i$ in $X_i$. Denote the data matrix for the adjusted strategies by $X^{(i)}_{0,1}$. For the remaining strategies in $X_0$, adjust them so they have zero mean in-sample (denote the corresponding data matrix by $X^{(i)}_{0,0}$). Arrange $X^{(i)}_{0,1}$ and $X^{(i)}_{0,0}$ into a new data matrix $Y_i$.
\item[Step III.] Bootstrap the time periods $J$ times. For each bootstrapped sample, calculate the realized error rates (or odds ratio) for $Y_i$, denoted by $f_{i,j}$ ($f$ stands for a generic error rate that is a function of the testing outcomes).
\item[Step IV.] \indent Repeat Steps I to III $I$ times. Calculate the final bootstrapped error rate as \newline $\frac{1}{IJ}\sum_{i=1}^I\sum_{j=1}^J f_{i,j}$.
\end{enumerate}

Again, keep in mind that the calculation of the realized error rate (i.e., $f_{i,j}$) in Step III requires the specification of the significance threshold (or a particular data-dependent testing procedure). As a result, the final bootstrapped error rates produced by our model are (implicitly) functions of the threshold.

\subsection{Type II Error Rates under Multiple Tests}
\indent While our definition of the Type I error rate (i.e., the FDR) is intuitive and fairly standard in the finance and statistics literatures, our definition of the Type II error rate (i.e., the false omission rate) deserves further discussion.

First, the way we define the Type II error rate is analogous to how we define the FDR. This can be seen by noting that while the FDR uses the total number of positives (i.e., \emph{TP} + \emph{FP}) as the denominator, our Type II error rate uses the total number of negatives (i.e., \emph{TN} + \emph{FN}). Several papers in statistics also recommend using the false omission rate --- also referred to as the false non discovery rate --- to measure the Type II error rate in the context of multiple testing (e.g., Genovese and Wasserman (2002), Sarkar (2006)).

The usual definition of the Type II error rate under single testing translates into the expected fraction of false negatives out of the total number of alternatives (i.e., the false negative rate, $\frac{FN}{FN+TP}$) in multiple testing. While we can easily accommodate this particular definition, we believe that the false omission rate is more appropriate under multiple testing and more suited to situations that are relevant to finance.


Note that these alternative definitions of the Type II error rate are transformations of each other and the Type I error rate, so we do not lose information by focusing on the false omission rate. For a given data set, the number of alternatives (i.e., $FN + TP$) and the number of nulls (i.e., $TN + FP$) are both fixed. There are only two unknowns among the four numbers. So any Type II error rate definition will be implied by the Type I error rate and the false omission rate.\footnote{To be precise, the realized error rates are explicit functions of the two unknowns. So this is only approximately true after taking expectations.}

The Type II error rate is linked to power. As a result, our Type II error rate implies a different interpretation of power that is best demonstrated with an example. Suppose that there are 100 managers, of which five are skilled. Suppose further that our test procedure correctly declares all 95 unskilled managers as unskilled and identifies three of the five skilled managers as skilled. The Type II error could be defined as 2/5, implying that 60\% ($=1-2/5$) of managers are correctly identified, which corresponds to the usual definition of power in a single test. Now suppose that we increase the total number of managers to 1,000 but we have the same number of skilled managers, five, and the same number of identified skilled managers, three. This would imply the same Type II error rate as before (i.e., 2/5), making the two cases indistinguishable from a Type II error perspective. However, it seems far more impressive for a testing method to correctly identify three out of five skilled managers among 1,000 managers than among 100. Our definition of the Type II error rate gets at exactly this difference --- we obtain an error rate of 2/97 ($=2/(95+2)$) for 100 tests versus 2/997 ($=2/(995+2)$) for 1,000 tests.

While we focus on the false omission rate in defining the Type II error rate, our bootstrap-based approach can easily accommodate alternative definitions such as the false negative rate or even definitions that have differential weights on individual errors. We view this as an important advantage of our framework in comparison to those provided by existing statistics literature (e.g., Genovese and Wasserman (2002), Sarkar (2006)).

\subsection{Bootstrapped Error Rates for Fama and French (2010)}
\indent Fama and French (2010) focus on the overall null hypothesis of zero alpha across mutual funds to test a version of the efficient market hypothesis. Therefore, the relevant error rates in their context are the probability of rejecting this overall null hypothesis when it is true (Type I error) and the probability of not rejecting this null hypothesis when some funds have the ability to generate a positive alpha (Type II error). We apply our framework described in Section I.B and I.C to find the corresponding Type I (denoted by $TYPE1_{ff}$) and Type II (denoted by $TYPE2_{ff}$) error rates.

We first focus on the Type I error. This corresponds to the case of $p_0 = 0$ in our framework. Let the original data be $X_0$. Similar to Fama and French (2010), we subtract the in-sample alpha estimate from each fund's return to generate a ``pseudo" sample of funds whose alphas are precisely zero. Let this sample be $Y_0$. We treat $Y_0$ as the population of fund returns for which the overall null hypothesis of zero alpha across all funds is true.

Fama and French (2010) bootstrap $Y_0$ to generate distributions of the cross-section of $t$-statistics and compare these bootstrapped distributions to the actual distribution to draw inferences. Similar to their strategy (and that in KTWW), if $Y_0$ can be treated as the ``pseudo" sample of fund returns under the null hypothesis, then any bootstrapped version of $Y_0$ can also be regarded as a ``pseudo" sample. This provides the basis for our evaluation of the Type I error rate for Fama and French (2010). We bootstrap $Y_0$ many times to generate alternative samples for which the overall null hypothesis of zero alpha across funds is true. For each sample, we apply Fama and French (2010) to see if the null hypothesis is (falsely) rejected. We take the average rejection rate as the Type I error rate.
%

More specifically, we bootstrap the time periods to perturb $Y_0$. This is where our approach departs from Fama and French (2010) --- while they make a one-shot decision for $Y_0$, we employ perturbations of $Y_0$ to simulate the error rates of the Fama-French (2010) approach. Let the perturbed data be denoted by $Y_i$. Notice that due to sampling uncertainty, fund alphas are no longer zero for $Y_i$, although the overall null hypothesis is still true since $Y_i$ is obtained by simply perturbing $Y_0$. For $Y_i$, we perform the Fama-French (2010) approach and let the testing outcome be given by $h_i$, where $h_i$ is equal to one if we reject the null hypothesis and zero otherwise. Finally, we perturb $Y_0$ many times and calculate the empirical Type I error rate (i.e., $TYPE1_{ff}$) as the average $h_i$.

We follow a procedure that is similar to that described in Section I.C to generate the Type II error rate. In particular, for a given fraction $p_0$ of funds with a positive alpha, we bootstrap the time periods and identify the top $p_0$ of funds with the highest $t$-statistics for alpha. We find the corresponding funds in the original data and adjust them such that they have the same alphas as those for the top $p_0$ of funds in the bootstrapped sample (denote the data matrix for the adjusted strategies by $X^{(i)}_{1,0}$). At the same time, we adjust the returns of the remaining funds so that they have zero alpha (denote the corresponding data matrix by $X^{(i)}_{0,0}$). By joining $X^{(i)}_{1,0}$ with $X^{(i)}_{0,0}$, we obtain a new panel. Using this panel, we apply the Fama-French (2010) approach and record the testing outcome as $l_i=1$ if the null hypothesis is not rejected (and zero otherwise). The empirical Type II error rate (i.e., $TYPE2_{ff}$) is then calculated as the average $l_i$.

In summary, while Fama and French (2010) make a one-time decision about whether to reject the overall null hypothesis for a given set of fund returns, our approach allows us to calibrate the error rates of Fama and French (2010) by repeatedly applying their method to bootstrapped data that are generated from the original data.

\subsection{Discussion}
\indent While traditional frequentist single-hypothesis testing focuses on the Type I error rate, more powerful frequentist approaches (e.g., likelihood-ratio test) and Bayesian hypothesis testing account for both Type I and Type II error rates. However, as we mention above, in the context of multiple testing, it is difficult to evaluate the Type II error rate for at least two reasons. First, the large number of alternative hypotheses makes such evaluation a potentially intractable multidimensional problem. Second, the multidimensional nature of the data, in particular the dependence across tests, confounds inference on the joint distribution of error rates across tests. Our framework provides solutions to both of these problems.

Given the large number of alternative hypotheses, it seems inappropriate to focus on any particular parameter vector as the alternative hypothesis. Our double-bootstrap approach simplifies the specification of alternative hypotheses by grouping similar alternative hypotheses. In particular, all alternative hypotheses that correspond to the case in which a fraction $p_0$ of the hypotheses are true (and the rest are false and hence set at the null hypotheses) are grouped under a single $H_1$ that is associated with $p_0$.

While our grouping procedure helps reduce the dimension of the alternative hypothesis space, a large number of alternative hypotheses associated with a given $p_0$ exist. However, we argue that many of these alternative hypotheses are unlikely to be true for the underlying data-generating process and hence irrelevant for hypothesis testing. First, hypotheses with a low in-sample $t$-statistic are less likely to be true than hypotheses with a high in-sample $t$-statistic. Second, the true parameter value for each effect under consideration is not arbitrary, but rather can be estimated from the data. Our bootstrap-based approach takes both of these points into account. In particular, our ranking of $t$-statistics based on the bootstrapped sample in Step I addresses the first issue, and our assignment of the actual (observed returns) as the ``true" parameter values to a fraction $p_0$ of tests that are deemed to be true after Step I addresses the second issue.

After we fix $p_0$ and find a parameter vector that is consistent with $p_0$ via the first-stage bootstrap, our second-stage bootstrap allows us to evaluate any function of error rates (both Type I and Type II) nonparametrically. Importantly, bootstrapping provides a convenient method to take data dependence into account, as argued in Fama and French (2010).

Our framework also allows us to make data-specific recommendations for the statistical cutoffs, and to evaluate the performance of any particular multiple-testing adjustment. In our applications, we provide examples of both. Given the large number of multiple testing methods that are available and the potential variation in a given method's performance across data sets, we view it important to generate data-specific statistical cutoffs that achieve a pre-specified Type I or Type II error constraint (or a combination of both). For example, Harvey, Liu, and Zhu (2016) apply several well-known multiple testing methods to anomaly data to control for the FDR. However, whether these or other methods can achieve the pre-specified FDR for these anomaly data is not known. Our double-bootstrap approach allows us to accurately calculate the FDR conditional on the $t$-statistic cutoff and $p_0$. We can therefore choose the $t$-statistic cutoff that exactly achieves a pre-specified FDR for values of $p_0$ that are deemed plausible.

While our paper focuses on estimating statistical objectives such as the FDR, one may attempt to apply our approach to economic objectives such as the Sharpe ratio of a portfolio of strategies or funds. However, there are several reasons to caution against such an attempt.

First, there seems to be a disconnect between statistical objectives and economic objectives in the finance literature. This tension traces back to two strands of literature. One is the performance evaluation literature, where researchers or practitioners, facing thousands of funds to select from, focus on simple statistics (i.e., statistical objectives) such as the FDR. The other strand is the mean-variance literature, where researchers record combinations of assets to generate efficient portfolios. Our framework is more applicable to the performance evaluation literature.\footnote{Note that although models used in the performance evaluation literature can generate a risk-adjusted alpha that indicates a utility gain (under certain assumptions of the utility function) by allocating to a particular fund, they still do not account for the cross-sectional correlations in idiosyncratic fund returns that impact allocation across a group of funds, which is the objective of the mean-variance literature.} Following the logic of this literature, we are more interested in estimating economically motivated quantities such as the fraction of outperforming funds than trying to find optimal combinations of funds due to either practical constraints (e.g., difficult to allocate to many funds) or computational issues (i.e., the cross-section is too large to reliably estimate a covariance matrix). While certain extensions of conventional statistical objectives in multiple testing (e.g., differential weights on the two types of error rates) are allowed in our framework, these extensions should still be considered rudimentary from the standpoint of economic objectives.


Second, statistical objectives and economic objectives may not be consistent with each other. For example, an individual asset with a high Sharpe ratio, which would likely be declared significant in a multiple-testing framework, may not be as attractive in the mean-variance framework once its correlations with other assets are taken into account.\footnote{Here correlations may be caused by correlations in idiosyncratic risks that are orthogonal to systematic risk factors.} As another example, the quest for test power may not perfectly align with economic objectives as methods that have low statistical power may still have important implications from an economic perspective (Kandel and Stambaugh (1996)).


We require that $p_0$ be relatively small so that the top $p_0\times N$ strategies in the first-stage bootstrap always have positive means.\footnote{This is the case for our application since we test the one-sided hypothesis for which the alternative hypothesis is that the strategy mean is positive. We do not require such a restriction on $p_0$ if the hypothesis test is two-sided.} This ensures that the population means for strategies that are deemed to be true in the second-stage bootstrap are positive, which is consistent with the alternative hypotheses and is required by our bootstrap approach. This is slightly stronger than requiring that $p_0$ be smaller than the fraction of strategies that have positive means in the original sample, since we need this (i.e., $p_0$ is smaller than the fraction of strategies with positive means) to be the case for all bootstrapped samples.\footnote{While imposing the constraint that $p_0$ is less than the fraction of strategies with a positive mean in the original sample cannot in theory rule out bootstrapped samples for which this condition is violated, we find zero occurrences of such violations in our results since $p_0$ is set to be below the fraction of strategies with a positive mean by a margin. One reason that we believe helps prevent such violations is the stability of order statistics, which ensures that the fraction of strategies with a positive mean in bootstrapped samples is fairly close to that in the original sample.} In our applications, we ensure that our choices of $p_0$ always meet this condition. We believe our restriction on $p_0$ is reasonable, as it is unlikely that every strategy with a positive return is true --- due to sampling uncertainty, some zero-mean strategies will have positive returns.


\section{Applications}
\indent We consider two applications of our framework. In the first, we study two large groups of investment strategies. We illustrate how our method can be used to select outperforming strategies and we compare our method to existing multiple-testing techniques. Our second application revisits Fama and French's (2010) conclusion about mutual fund performance. We show that Fama and French's (2010) technique is underpowered in that it is unable to identify truly outperforming funds. Overall, our two applications highlight the flexibility of our approach in addressing questions related to the multiple-testing problem.

\subsection{Investment Strategies}
\subsubsection{Data Description: CAPIQ and 18,000 Anomalies}
\indent We start with Standard and Poor's Capital IQ (CAPIQ) database, which covers a broad set of ``alpha" strategies. This database records the historical performance of synthetic long-short strategies that are classified into eight groups based on the types of their risk exposures (e.g., market risk) or the nature of their forecasting variables (e.g., firm characteristics). Many well-known investment strategies are covered by this database, for example, CAPM beta (Capital Efficiency Group), value (Valuation Group), and momentum (Momentum Group). For our purposes, we study 484 strategies (i.e., long-short strategies) for the U.S. equity market from 1985 to 2014.\footnote{We thank CAPIQ for making these data available to us. Harvey and Liu (2017) also use the CAPIQ data to study factor selection.}

The CAPIQ data contain a large number of ``significant" investment strategies based on single-hypothesis tests, which distinguishes it from the other data sets that we consider below (i.e., anomaly data or mutual funds). This is not surprising as CAPIQ is biased towards a select group of strategies that are known to historically perform well. As such, it is best to treat these data as an illustration of how our method helps address concerns about Type II errors. One caveat in using the CAPIQ data (or other data that include a collection of investment strategies, such as Hou, Xue, and Zhang (2020)) relates to the selection bias in the data, which is due mainly to publication bias. We do not attempt to address publication bias in this paper. See Harvey, Liu, and Zhu (2016) for a parametric approach to address publication bias.

The second data set comprises the 18,113 anomalies studied in Yan and Zheng (2017). Yan and Zheng (2017) construct a comprehensive set of anomalies based on firm characteristics.\footnote{We thank Sterling Yan for providing us their data on anomaly returns. } Using the Fama and French (2010) approach to adjust for multiple testing, they claim that a large portion of the anomalies are true and argue that there is widespread mispricing. We revisit the inference problem faced by Yan and Zheng (2017). Our goal is to use our framework to calibrate the error rates of their approach and offer insights on how many anomalies are true in their data --- and we reach a different conclusion.

Overall, the two data sets represent two extreme cases for a pool of investment strategies that researchers or investors may seek to analyze. While the CAPIQ data focus on a select set of backtested strategies, the 18,000 anomalies include a large collection of primitive strategies obtained from a data mining exercise.

\subsubsection{Preliminary Data Analysis}
\indent We use the $t$-statistic to measure the statistical significance of an investment strategy. For the CAPIQ data, we simply use the $t$-statistic of the strategy return.\footnote{Our results for the CAPIQ data are similar if we adjust strategy returns using, say, the Fama-French-Carhart four-factor model.} For the 18,113 anomalies in Yan and Zheng (2017), we follow their procedure and calculate excess returns with respect to the Fama-French-Carhart four-factor model.\footnote{Yan and Zheng (2017) use several specifications of the benchmark model, including the Fama-French-Carhart four-factor model. We focus only on the Fama-French-Carhart four-factor model to save space as well as to illustrate how benchmark models can be easily incorporated into our framework. Our results are qualitatively similar if we use alternative benchmark models.}
Yan and Zheng (2017) construct their strategies using firm characteristics and these strategies likely have high exposures to existing factors. While we focus on simple $t$-statistics to describe our framework in the previous section, our bootstrap-based method can be easily adapted to incorporate any benchmark factors.\footnote{To preserve cross-sectional correlations, factors need to be resampled simultaneously with strategy returns when we resample the time periods.}

\begin{figure}[ht]
\vspace*{-2cm}
\centering
\includegraphics[trim=1.0cm 4.5cm 0cm 4.0cm, clip= true,width= 1.1\textwidth]{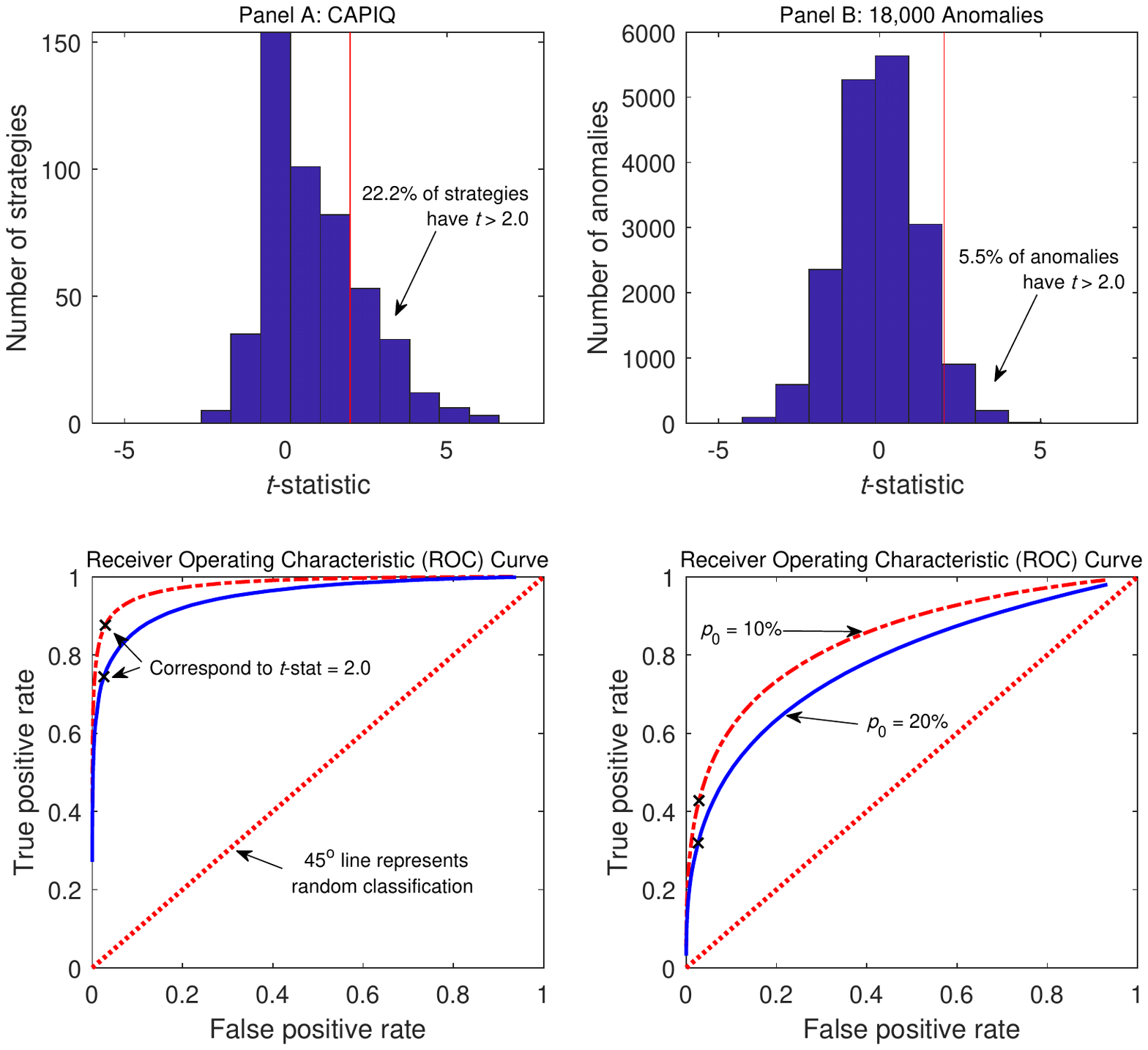}
\vspace*{-2cm}
\caption{\textbf{Preliminary data analysis.} \textmd{\small{This figure shows $t$-statistic distributions and receiver operating characteristic (ROC) curves for CAPIQ and 18,000 anomalies.} We plot the $t$-statistic distributions (the top figures in Panel A and B) for the 484 investment strategies in the CAPIQ data and the 18,113 strategies in Yan and Zheng (2017). The $t$-statistic is given as the $t$-statistic for the original strategy return for the CAPIQ data and the $t$-statistic for the Fama-French-Carhart four-factor model-adjusted anomaly alpha for the 18,113 strategies. The bottom figures in Panel A and B show the ROC curves, corresponding to $p_0 = 10\%$ and 20\%. The crosses on the ROC curves mark the points that correspond to a $t$-statistic cutoff of 2.0.}}
\label{fig:ROC}
\end{figure}

In Figure \ref{fig:ROC}, the top graph in Panel A (Panel B) shows the distribution  of the $t$-statistics for the CAPIQ (18,000 anomalies) data. For the CAPIQ data, the distribution is skewed to the right. The fraction of strategies that have a $t$-statistic above 2.0 is 22.1\% (=107/484). For the 18,000 anomalies, the distribution is roughly symmetric around zero and the fraction of strategies that have a $t$-statistic above 2.0 is
 5.5\% (=989/18,113). These statistics are consistent with how these two data sets are assembled.

Notice that the direction of the long-short strategies for the 18,113 anomalies data is essentially randomly assigned, which explains the symmetry of the distribution of the $t$-statistics. This also suggests that we should perform two-sided hypothesis tests, as both significant outperformance and significant underperformance (relative to benchmark factor returns) qualify as evidence of anomalous returns. However, to be consistent, and thus facilitate comparison with our results for CAPIQ, we illustrate our method via one-sided tests and test only for outperformance. Our method can be readily applied to the case of two-sided tests. Moreover, given the data-mining nature of the 18,113 anomalies data (so the two tails of the distribution of the $t$-statistics for anomalies are roughly symmetric), the Type I (Type II) error rates for one-sided tests under $p_0$ are approximately one-half of the Type I (Type II) error rates under $2p_0$. While we present results only for one-sided tests, readers interested in two-sided tests can apply the above transformation to obtain the corresponding error rates for two-sided tests.


We illustrate how our method can be used to create the receiver operating characteristic (ROC) curve, which is an intuitive diagnostic plot to assess the performance of a classification method (e.g., a multiple-testing method).\footnote{See Fawcett (2006) and Hastie, Tibshirani and Friedman (2009) for applications of the ROC method.} It plots the true positive rate (TPR), defined as the number of true discoveries over the total number of true strategies, against the false positive rate (FPR), defined as the ratio of false discoveries to the total number of zero-mean strategies.


Our framework allows us to use bootstrapped simulations to draw the ROC curve. In particular, for a given $p_0$, the first-round bootstrap of our method classifies all strategies into true strategies and zero-mean strategies. The second-round bootstrap then calculates the realized TPR and FPR for each $t$-statistic cutoff for each bootstrapped simulation. We simulate many times to generate the average TPR and FPR across simulations.

Note that our previously defined FDR is different from FPR. Although the numerator is the same (i.e., the number of false discoveries), the denominator is the total number of true strategies (FPR) versus the total number of discoveries (FDR).\footnote{Conceptually, FDR is a more stringent error rate definition than FPR when there are a large number of false strategies and the signal-to-noise ratio is low in the data. In this case, a high $t$-statistic cutoff generates very few discoveries. But FDR could still be high since it is difficult to distinguish the good from the bad. In contrast, since the denominator for FPR is larger than that for FDR, FPR tends to be much lower than FDR.} Our framework can flexibly accommodate alternative error rate definitions that are deemed useful.

Figure \ref{fig:ROC} also shows the ROC curve for $p_0 = 10\%$ and 20\%. On the ROC graph, the ideal classification outcome is given by the point $(0,1)$, that is, $FPR = 0$ and $TPR = 100\%$. As a result, a ROC curve that is closer to $(0,1)$ (or further away from the 45-degree line which corresponds to random classification) is deemed better. Two patterns can be seen in Figure \ref{fig:ROC}. First, the ROC curve is better for the CAPIQ data than for the 18,000 anomalies. Second, for each data set, a smaller $p_0$ results in a better ROC curve.

These two patterns reflect key features of the two data sets that we analyze. The better ROC curve for the CAPIQ data stems from the higher average signal-to-noise ratio in these data as the CAPIQ contain a large number of select investment strategies. Although the higher signal-to-noise ratio can also be seen from the distribution of $t$-statistics (i.e., the top graph of Figure \ref{fig:ROC}), the ROC curve quantifies this through FPR and TPR, which are potentially informative metrics in classifying investment strategies.\footnote{Another benefit is that the ROC curve generated in our framework takes test correlations into account (since, similar to Fama and French (2010), our second-stage bootstrap generates the same resampled time periods across all strategies), whereas the $t$-statistic distribution cannot.} However, a smaller $p_0$ results in a more select group of strategies (e.g., the average $t$-statistic is higher for a smaller $p_0$), which explains the better classification outcome as illustrated by the ROC curve. The ROC curve highlights the trade-off between FPR and TPR for different levels of $p_0$.\footnote{Note that it is straightforward to find the optimal FPR (and the corresponding $t$-statistic cutoff) associated with a certain trade-off between FPR and TPR. For example, if we equally weight the FPR and the TPR (i.e., we try to maximize TPR $-$ FPR), then the optimal FPR is given by the tangency point of a 45-degree tangent line to the ROC curve. }

%
%
%

\subsubsection{The Selection of Investment Strategies: Having a Prior on $p_0$.}
\indent We first apply our method to study how the Type I and Type II error rates vary across different $t$-statistic thresholds. In practice, researchers often use a pre-determined $t$-statistic threshold to perform hypothesis testing, such as 2.0 at the 5\% significance level for a single hypothesis test, or a $t$-statistic that exceeds 3.0 based on Harvey, Liu, and Zhu (2016).\footnote{Notice that it was never the intention of Harvey, Liu and Zhu (2016) to recommend the 3.0 threshold as a universal rule that applies to any data set. The purpose of the current paper is to show how one can use our method to calibrate Type I and Type II error rates based on different $t$-statistic thresholds, through which one can obtain the ``optimal" $t$-statistic threshold that applies to the particular data set at hand (see also Harvey (2017)).} We investigate the implications of these choices using our method.

\begin{figure}[H]

\hspace*{-1cm}
\centering
\includegraphics[trim=1.5cm 4cm 0cm 6.0cm, clip= true,width= 1.2\textwidth, height = 1.2\textwidth]{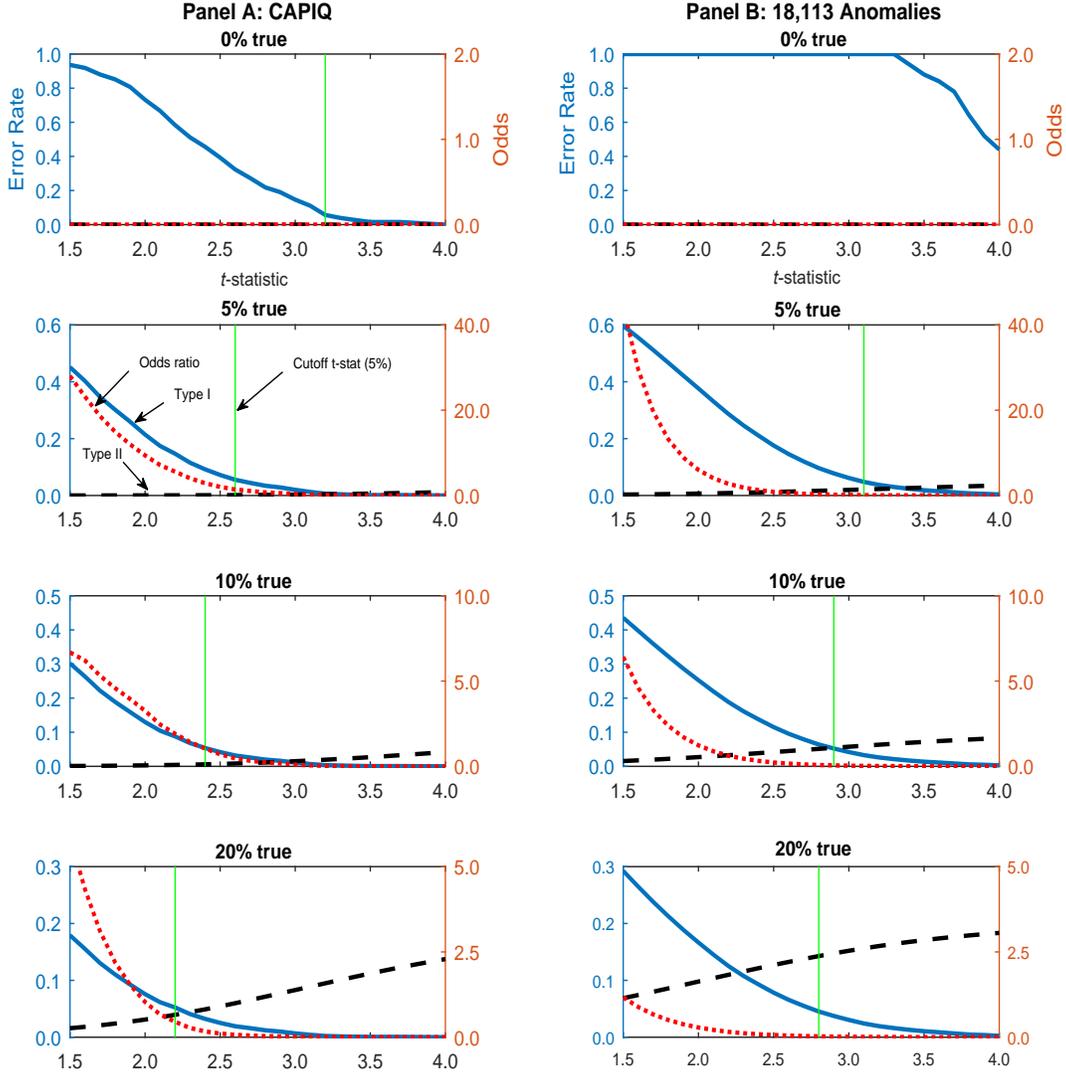}
\vspace*{-4cm}
\caption{\textbf{Error rates for fixed $t$-statistic thresholds: CAPIQ and 18,113 anomalies.} \textmd{\footnotesize{This figure shows simulated Type I and Type II error rates for CAPIQ and the 18,113 anomalies.} For each fraction $p_0$ of strategies that are believed to be true, we follow our method in Section 2 and set $I = 100$ (for each $i$, we bootstrap to obtain the ranking of strategies and set the top $p_0$ as true) and $J = 1,000$ (conditional on $i$, for each $j$, we bootstrap the time periods) to run 100,000 ($=100\times 1,000$) bootstrapped simulations to calculate the empirical Type I (fraction of false discoveries) and Type II (fraction of missed discoveries among all non discoveries) error rates. On the graph, the (blue) solid line is the Type I error rate, the (black) dashed line is the Type II error rate, and the (red) dotted line is the odds ratio. The left axis is the error rate that applies to the Type I and Type II error rates, whereas the right axis is the odds ratio. The vertical (green) line marks the $t$-statistic cutoff that corresponds to a Type I error rate of 5\%. The $t$-statistic cutoff for the top figure in Panel B (4.9) is omitted.}}
\label{fig:fix_IQ}
\end{figure}

Figure \ref{fig:fix_IQ} shows the error rates across a range of $t$-statistics for both the CAPIQ data and the 18,000 anomalies. We see that the classic trade-off between Type I and Type II error rates in single hypothesis tests continues to hold in a multiple testing framework. When the threshold $t$-statistic increases, the Type I error rate (the rate of false discoveries among all discoveries) declines while the Type II error rate (the rate of misses among all non discoveries) increases. In addition, the odds ratio, which is the ratio of false discoveries to misses, also decreases as the threshold $t$-statistic increases. We highlight the threshold $t$-statistic that achieves 5\% significance in Figure \ref{fig:fix_IQ}.\footnote{The threshold $t$-statistic is 4.9 for the 18,000 anomalies when $p_0=0$. Since we set the range of the $t$-statistic (i.e., x-axis) to span from 1.5 to 4.0, we do not display this $t$-statistic in Figure \ref{fig:fix_IQ}.} We see that this threshold $t$-statistic decreases as $p_0$ (the prior fraction of true strategies) increases. This makes sense as a higher prior fraction of true strategies calls for a more lenient $t$-statistic threshold.

How should an investment manager pick strategies based on Figure \ref{fig:fix_IQ}? First, the manager needs to elicit a prior on $p_0$, which is likely driven by both her previous experience and the data (e.g., the histogram of $t$-statistics that we provide in Section II.A.1). Suppose that the manager believes that a $p_0$ of 10\% is plausible for the CAPIQ data. If she wants to control the Type I error rate at 5\%, she should set the $t$-statistic threshold to $t=2.4$ (see Panel A of Figure \ref{fig:fix_IQ}, $p_0=10\%$). Based on this $t$-statistic threshold, 18\% of strategies survive based on the original data.

Alternatively, under the same belief on $p_0$, suppose that the investment manager is more interested in balancing Type I and Type II errors and wants to achieve an odds ratio of around $1/5$, that is, on average five misses for each false discovery (alternatively, she believes that the cost of a Type I error is five times that of a Type II error). Then she should set the $t$-statistic threshold to $t=2.6$ if she is interested in $p_0 = 10\%$ and 15\% of the strategies survive.


Therefore, under either the Type I error rate or the odds ratio, neither 2.0 (the usual cutoff for 5\% significance) nor 3.0 (based on Harvey, Liu, and Zhu (2016)) is optimal from the perspective of the investor. The preferred choices lie between 2.0 and 3.0 for the examples we consider and depend on both $p_0$ and the CAPIQ data that we study.

Comparing Panels A and B in Figure \ref{fig:fix_IQ}, at $p_0 = 0$, the much higher $t$-statistic cutoff for the 18,113 anomalies (i.e., 4.9) than that for CAPIQ (i.e., 3.2) reflects the larger number of strategies for the anomaly data (thresholds need to be higher when more strategies are tested, that is, when the multiple testing problem is more severe). For alternative values of $p_0$, the higher $t$-statistic cutoff for the anomaly data is driven by its low signal-to-noise ratio, which is also seen from the analysis of the ROC curve.

The overall message of Figure \ref{fig:fix_IQ} is that two elements need to be taken into account to select the best strategies. One is investors' prior on the fraction of true strategies. Most economic agents are subjective Bayesians, so it is natural to try to incorporate insights from the Bayesian approach into a multiple-testing framework.\footnote{See Harvey (2017) for the application of the Bayesian approach to hypothesis testing.} Our method allows us to evaluate the sensitivity of the error rates to changes in the prior fraction of true strategies. Instead of trying to control the error rates under all circumstances as in the traditional frequentist hypothesis testing framework, our view is that we should incorporate our prior beliefs into decision-making process, as doing so helps us make informed decisions based on the trade-off between the Type I and Type II errors.

The other element needed to select the best strategies relates to the particular data at hand and the hypothesis that is being tested. While many traditional multiple-testing adjustments work under general assumptions on the dependence structure in the data, they may be too conservative in that they underreject when the null hypothesis is false, leading to too many Type II errors (as we shall see below). Our bootstrap-based framework generates $t$-statistic thresholds that are calibrated to the particular decision being considered and the particular data under analysis.

\begin{figure}[ht]
\vspace*{-1cm}
\centering
\includegraphics[trim=2.0cm 4.5cm 0cm 4.0cm, clip= true,width= 0.8\textwidth, height= 0.6\textheight]{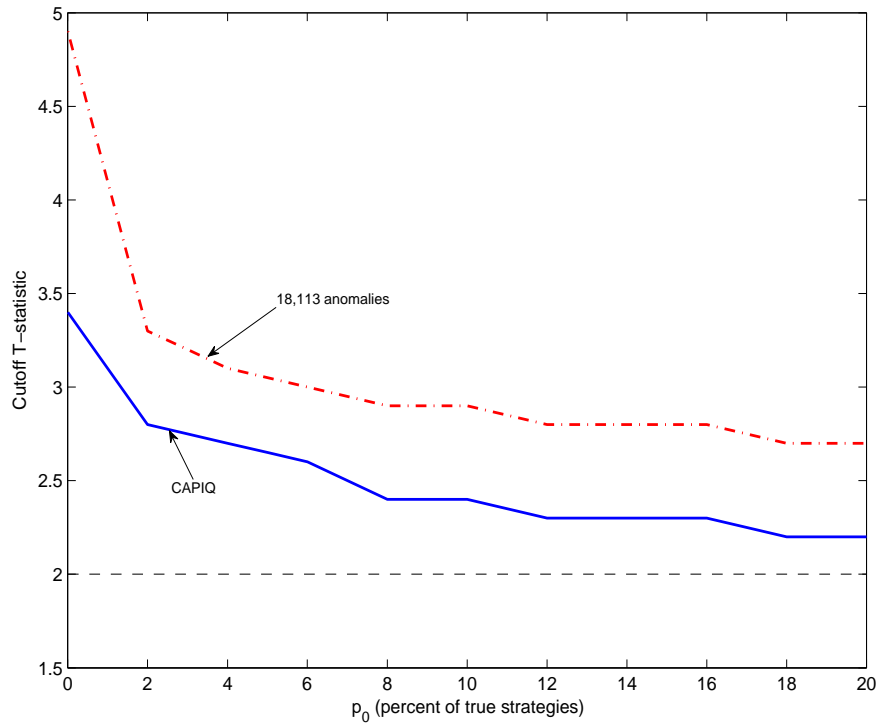}
\vspace*{-1cm}
\caption{\textbf{Cutoff $t$-statistics as a function of $p_0$ for CAPIQ and 18,113 anomalies.} \textmd{\small{This figure shows cutoff $t$-statistics as a function of $p_0$ for CAPIQ investment strategies and 18,113 anomalies. For each hypothetical level of $p_0$ between 0\% and 20\% (with 1\% increments), we search for the optimal cutoff $t$-statistic between 1.5 and 5.0 (with 0.1 increments) that corresponds to a Type I error rate of 5\%. The kinks in the graph are caused by the discrete increments in both $p_0$ and the range of $t$-statistics from which we search.}}}
\label{fig:IQ_ts}
\end{figure}

Figure \ref{fig:IQ_ts} plots the cutoff $t$-statistics that generate a Type I error rate of 5\% against $p_0$. In general, the cutoff $t$-statistic declines as $p_0$ becomes larger, since a discovery is less likely to be false when a larger fraction of strategies are true.\footnote{Our approach of choosing data-dependent cutoffs that guarantee particular Type I error rates is similar to the use of sample-size adjusted $p$-values in the econometrics literature. Both strategies aim to achieve a Type I error rate that is closer to the desired level, and thereby increase test power. Whereas sample-size adjusted $p$-values are usually applied to univariate tests from a time-series perspective, our approach applies to multiple testing and adjusts for the sample size (and distributional properties) for both the cross section and the time series.}

\subsubsection{The Selection of Investment Strategies: Unknown $p_0$}
\indent When $p_0$ is unknown, our framework can be used to evaluate the performance of multiple-testing corrections across different values of $p_0$. While multiple-testing methods are designed to control the FDR at the appropriate level regardless of the value of $p_0$, their performance in finite samples, particularly for the data under analysis, is unknown. Our method provides guidance on which method to use for a particular data set.

Given our focus on the FDR, we implement several popular multiple-testing adjustments proposed in the statistics literature that aim to control the FDR. Our goal is not to compare all existing methods. Rather, we focus on a few representative methods in terms of their theoretical properties and illustrate how to apply our framework to select the best method for the particular data being analyzed.

We want to emphasize that our data-driven approach is different from the usual simulation exercises carried out by studies that propose multiple-testing methods. Most simulation studies make simple assumptions about the underlying data structure (e.g., a multivariate normal distribution with a certain correlation structure).\footnote{See, for example, the simulation studies in Romano and Wolf (2005) and Romano, Shaikh, and Wolf (2008).} However, the data sets we typically encounter in financial economics (e.g., the cross-section of stock returns) have important features (e.g., missing data, cross-sectional dependence, tail dependence, etc.) that may make simplifying assumptions poor approximations. In our
multiple-testing context, many methods are known to be sensitive to some of these features (i.e., cross-sectional dependence and tail dependence). It is therefore important to customize the performance evaluation of a multiple-testing method to the data being examined.

The first set of multiple-testing adjustments that we consider are the Benjamini and Hochberg (1995) adjustment, BH, which controls the FDR under the pre-specified value if tests are mutually independent, and Benjamini and Yekutieli (2001) adjustment, BY, which controls the FDR under arbitrary dependence of the tests.\footnote{Benjamini and Yekutieli (2001) show that independence can be replaced by the weaker assumption of positive regression dependency as in Benjamini and Hochberg (1995).} BH may not work if tests are correlated in a certain fashion, while BY tends to be overly conservative (i.e., too few discoveries).\footnote{For further details on these two methods and other multiple-testing adjustments, as well as their applications in finance, see Harvey, Liu, and Zhu (2016) and Chordia, Goyal, and Saretto (2020).}

Given our focus on test power, we also consider the adjustment of Storey (2002), which sometimes results in an improvement over BH and BY in terms of test power. A plug-in parameter in Storey (2002), denoted by $\theta$, helps replace the total number of tests in BH and BY with an estimate of the fraction of true null hypotheses. Bajgrowicz and Scaillet (2012) suggest $\theta = 0.6$. We experiment with three values for $\theta$: $\theta = 0.4$, $0.6$, and $0.8$.

Finally, we consider a multiple-testing method that has strong theoretical properties in terms of the requirement on the data to achieve a pre-specified error rate.\footnote{We thank Laurent Barras for bringing this literature to our attention.} In particular, we implement the bootstrap-based approach of Romano, Shaikh, and Wolf (RSW, 2008), which controls the FDR asymptotically under arbitrary dependence in the data.\footnote{See Romano and Wolf (2005) for a companion approach that controls the FWER. } Our goal is to analyze the finite-sample performance of the RSW approach for the two data sets we study. However, implementation of RSW is computationally intensive (especially for the 18,000 anomalies) as it requires estimating $B\times O(M^2)$ regression models (where $B$ is the number of bootstrapped iterations and $M$ is the total number of tests) to derive a single $t$-statistic cutoff. Accordingly, we randomly sample from the 18,000 anomalies to reduce the sample size and apply RSW to the subsamples. But this makes our results not directly comparable with those based on the full sample. We therefore present our results using RSW in the Appendix.

In addition to the aforementioned methods, we report the Type II error rates derived under the assumption that $p_0$ is known ex ante and the $t$-statistic cutoff is fixed. In particular, for a given $p_0$, our previous analysis allows us to choose the $t$-statistic threshold such that the pre-specified Type I error rate is exactly achieved, thereby minimizing the Type II error rate of the test.\footnote{The Type II error rate is minimized in our framework in the following sense. Suppose we are only allowed to choose a fixed $t$-statistic cutoff for each assumed level of $p_0$. Imagine that we try to solve a constrained optimization problem where the pre-specified significance level is the Type I error rate constraint and our objective function seeks to minimize the Type II error rate. Given the trade-off between the Type I and Type II error rates, the optimal value for the objective function (i.e., the Type II error rate) is achieved when the constraint is met with equality.} This minimized Type II error rate is a useful benchmark to gauge the performance of other multiple-testing methods. Note that other multiple-testing methods may generate a smaller Type II error rate than the optimal rate because their Type I error rates may exceed the pre-specified levels.\footnote{There is also a difference in the Type I error rate between a fixed $t$-statistic cutoff and multiple-testing methods. Using our notation from Section I.B (i.e., $X_i$ denotes a particular parameterization of the hypotheses), while multiple-testing methods seek to control the FDR for each parameterization (i.e., $X_i$) of the hypotheses (i.e., $E(FDR|X_i)$), the fixed $t$-statistic cutoff that we use above aims to control $E(E(FDR|X_i)|p_0)$, which is the average $E(FDR|X_i)$ across different realizations of $X_i$ for a given $p_0$.}

Before presenting our results, we briefly report summary statistics on the correlations among test statistics for our data sets. For CAPIQ, the $10^{th}$ percentile, median, and $90^{th}$ percentile for the pairwise test correlations are -0.377, 0.081, and 0.525; the corresponding absolute values of the pairwise correlations are 0.040, 0.234, and 0.584. For the 18,000 anomalies, these numbers are -0.132, 0.003, and 0.145 (original values) and 0.013, 0.069, and 0.186 (absolute values). While correlations should provide some information on the relative performance of different methods, other features of the data may also be important, as we shall see below.

Tables \ref{table:exist_IQ} and \ref{table:exist_YanZheng} report results for the CAPIQ and the 18,000 anomalies, respectively. In Table \ref{table:exist_IQ}, BH generates FDRs that are always below and oftentimes close to the pre-specified significance levels, suggesting overall good performance. In comparison, BY is too conservative in that it generates too few discoveries, resulting in higher Type II error rates compared to BH. The three Storey tests generally fail to meet the pre-specified significance levels, although the Type I error rates generated by Storey ($\theta = 0.4$) are reasonably close. Overall, BH seems to be the preferred choice for CAPIQ among the five tests we examine.
Storey ($\theta$ = 0.4) is also acceptable, although one has to be cautious about the somewhat higher Type I error rates than the desired levels.

The test statistic from our method (i.e., the last column in Table \ref{table:exist_IQ}) provides gains in test power compared to BH, BY, and Storey. In particular, compared to other methods that also meet the pre-specified significance levels (i.e., BH and BY as in Table \ref{table:exist_IQ}), the Type II error rate for this test statistic is uniformly lower. For example, for $p_0=20\%$ and a significance level of 10\%, all three Storey tests deliver a Type I error that is too high, at 10\%. When we perform power comparison, we omit the Storey methods and focus on the two other models (i.e., BH and BY). Between these two, BH is the better-performing model since its Type I error rate is closer to the pre-specified significance level. We therefore use it as the benchmark to evaluate the power improvement. Compared with BH, which generates a Type II error rate of 3.5\%, our model produces an error rate of 2.9\%.

For the 18,113 anomalies, the results in Table \ref{table:exist_YanZheng} present a different story than Table \ref{table:exist_IQ}. In particular, BH generally performs worse than the three Storey tests in controlling the FDR when $p_0$ is not too large (i.e., $p_0 \leq 10\%$), although the Storey tests also lead to higher Type I error rates than desired. Overall, BY should be the preferred choice to strictly control the Type I error rate at the pre-specified level, although it appears to be far too conservative. Alternatively, Storey ($\theta=0.8$) dominates the other methods in achieving the pre-specified significance levels when $p_0 \leq 10\%$.

The Appendix presents results on the finite-sample performance of RSW applied to our data sets. Despite the strong theoretical appeal of RSW, it often leads to a higher Type I error rate than the desired level for both data sets. In fact, compared to the aforementioned multiple-testing methods we consider, RSW often generates the largest distortion in test size when $p_0$ is relatively small (i.e., $p_0 \leq 20\%$).

Our results highlight the data-driven nature of the performance of alternative multiple-testing methods. While BH works well theoretically when tests are independent, it is not clear how departures from independence affect its performance. Interestingly, our results show that BH performs much worse for the 18,000 anomalies than for CAPIQ, although the data for the 18,000 anomalies appear more ``independent" than CAPIQ based on the test correlations.\footnote{Note that we use the average correlation across strategies as an informal measure of the degree of cross-sectional dependence. However, correlation likely provides an insufficient characterization of dependence because certain forms of dependence may not be captured by the correlation. In our context, there may be important departures from independence in the 18,000 anomalies data that impact the performance of BH but are not reflected in the average correlation. Our results thus highlight the data-dependent nature of existing multiple-testing methods.} As another example, we show that Storey ($\theta = 0.4$) and Storey ($\theta = 0.8$) could be the preferred choice for the two data sets we examine, whereas $\theta = 0.6$ is the value recommended by Bajgrowicz and Scaillet (2012). Finally, methods that are guaranteed to perform well asymptotically may have poor finite-sample performance, as we show for RSW in the Appendix.

Overall, our method sheds light on which multiple-testing adjustment performs the best for a given data set. Of course, it is also possible to use our method directly to achieve a given false discovery level and to optimize the power across different assumptions for $p_0$.

%

\begin{landscape}
\begin{table}[H]
\centering 
  \footnotesize\addtolength{\tabcolsep}{-4pt}
  \caption{\textbf{Error Rates for Existing Methods: CAPIQ}  }
\captionsetup{width=22.0cm}
\caption*{\textmd{\small{The table presents simulated Type I and Type II error rates for CAPIQ data.} For each fraction $p_0$ of strategies that are believed to be true, we follow our method in Section I and set $I = 100$ (for each $i$, we bootstrap to obtain the ranking of strategies and set the top $p_0$ as true) and $J = 1,000$ (conditional on $i$, for each $j$, we bootstrap the time periods) to run 100,000 ($=100\times 1,000$) bootstrapped simulations to calculate the empirical Type I and Type II error rates. For a given significance level $\alpha$, we set the Type I error rate at $\alpha$ and find the corresponding Type II error rate, which is the ``optimal" Type II error rate. We also implement BH (Benjamini and Hochberg (1995)), BY (Benjamini and Yekutieli (2001)), and Storey (2002) for our simulated data and calculate their respective Type I and Type II error rates. }}
\small{
\begin{tabular}{cccccccccccccccccccccccccc} 
\hline\hline
      && &&\multicolumn{9}{c}{Type I} &&& \multicolumn{11}{c}{Type II} \\
      \cline{5-13} \cline{16-26}
$p_0$ && $\alpha$ && $BH$ && $BY$  && \multicolumn{5}{c}{Storey} &&& $BH$ && $BY$  && \multicolumn{5}{c}{Storey} && $HL (opt)^*$ \\
      \cline{9-13}\cline{20-24}
      &&          &&      &&       &&($\theta = 0.4$)&& ($\theta=0.6$) && ($\theta = 0.8$) &&&       &&       &&($\theta = 0.4$)&& ($\theta=0.6$) && ($\theta = 0.8$) && \\
(frac. of true) && (sig. level) && && && && &&  &&& && && && && &&  \\
\hline
\multicolumn{14}{c}{} \\ [-3mm]
2\%   &&  1\%     && 0.010  &&  0.002 && 0.023 && 0.022 && 0.021 &&& 0.001 && 0.002 && 0.001 && 0.001 && 0.001 && 0.001 \\
      &&  5\%     && 0.044  &&  0.009 && 0.058 && 0.058 && 0.058 &&& 0.000 && 0.001 && 0.000 && 0.000 && 0.000 && 0.000  \\
      &&  10\%    && 0.086  &&  0.013 && 0.102 && 0.104 && 0.107 &&& 0.000 && 0.001 && 0.000 && 0.000 && 0.000 && 0.000    \\
      \multicolumn{14}{c}{} \\ [-3mm]
5\%   &&  1\%     && 0.008  &&  0.002 && 0.018 && 0.018 && 0.017 &&& 0.006 && 0.011 && 0.005 && 0.005 && 0.005 && 0.005 \\
      &&  5\%     && 0.047  &&  0.007 && 0.060 && 0.060 && 0.060 &&& 0.003 && 0.006 && 0.002 && 0.002 && 0.002 && 0.002  \\
      &&  10\%    && 0.091  &&  0.012 && 0.102 && 0.105 && 0.113 &&& 0.002 && 0.005 && 0.001 && 0.001 && 0.002 && 0.001    \\
      \multicolumn{14}{c}{} \\ [-3mm]
10\%   &&  1\%     && 0.008  &&  0.002 && 0.017 && 0.017 && 0.016 &&& 0.022 && 0.037 && 0.021 && 0.021 && 0.020 && 0.021 \\
      &&  5\%     && 0.046  &&  0.006 && 0.058 && 0.059 && 0.063 &&& 0.011 && 0.024 && 0.010 && 0.010 && 0.010 &&  0.010 \\
      &&  10\%    && 0.087  &&  0.014 && 0.108 && 0.112 && 0.123 &&& 0.007 && 0.019 && 0.007 && 0.007 && 0.006 &&  0.007   \\
      \multicolumn{14}{c}{} \\ [-3mm]
20\%   &&  1\%     && 0.008 &&  0.002 && 0.016 && 0.016 && 0.016 &&& 0.082 && 0.117 && 0.088 && 0.077 && 0.074 &&  0.079 \\
      &&  5\%     && 0.042  &&  0.006 && 0.061 && 0.064 && 0.071 &&& 0.049 && 0.088 && 0.044 && 0.043 && 0.041 &&  0.047 \\
      &&  10\%    && 0.079  &&  0.012 && 0.117 && 0.123 && 0.142 &&& 0.035 && 0.074 && 0.030 && 0.029 && 0.028 &&  0.029   \\
            \multicolumn{14}{c}{} \\ [-3mm]
30\%   &&  1\%     && 0.007 &&  0.002 && 0.017 && 0.017 && 0.017 &&& 0.179 && 0.223 && 0.161 && 0.159 && 0.154 && 0.153  \\
      &&  5\%     && 0.037  &&  0.005 && 0.064 && 0.068 && 0.078 &&& 0.126 && 0.188 && 0.104 && 0.101 && 0.096 && 0.095  \\
      &&  10\%    && 0.069  &&  0.011 && 0.119 && 0.129 && 0.155 &&& 0.098 && 0.168 && 0.076 && 0.073 && 0.068 && 0.066    \\
\hline
\multicolumn{14}{l}{\footnotesize{*Type II calculated at optimized Type I error $= \alpha$}}
  \end{tabular}\par}
\label{table:exist_IQ} 
\end{table}
\end{landscape}

\begin{landscape}
\begin{table}[H]
\centering 
  \footnotesize\addtolength{\tabcolsep}{-4pt}
  \caption{\textbf{Error Rates for Existing Methods: 18,000 Anomalies}  }
\captionsetup{width=22.0cm}
\caption*{\textmd{\small{This table presents simulated Type I and Type II error rates for the 18,000 anomalies data.} For each fraction $p_0$ of strategies that are believed to be true, we follow our method in Section I and set $I = 100$ (for each $i$, we bootstrap to obtain the ranking of strategies and set the top $p_0$ as true) and $J = 1,000$ (conditional on $i$, for each $j$, we bootstrap the time periods) to run 100,000 ($=100\times 1,000$) bootstrapped simulations to calculate the empirical Type I and Type II error rates, as well as the odds ratio. For a given significance level $\alpha$, we set the Type I error rate at $\alpha$ and find the corresponding Type II error rate, which is the ``optimal" Type II error rate. We also implement BH (Benjamini and Hochberg (1995)), BY (Benjamini and Yekutieli (2001)), and Storey (2002) for our simulated data and calculate their respective Type I and Type II error rates.}  }
\small{
\begin{tabular}{cccccccccccccccccccccccccc} 
\hline\hline
      && &&\multicolumn{9}{c}{Type I} &&& \multicolumn{11}{c}{Type II} \\
      \cline{5-13} \cline{16-26}
$p_0$ && $\alpha$ && $BH$ && $BY$  && \multicolumn{5}{c}{Storey} &&& $BH$ && $BY$  && \multicolumn{5}{c}{Storey} && $HL (opt)^*$ \\
      \cline{9-13}\cline{20-24}
      &&          &&      &&       &&($\theta = 0.4$)&& ($\theta=0.6$) && ($\theta = 0.8$) &&&       &&       &&($\theta = 0.4$)&& ($\theta=0.6$) && ($\theta = 0.8$) && \\
(frac. of true) && (sig. level) && && && && &&  &&& && && && && &&  \\
\hline
\multicolumn{14}{c}{} \\ [-3mm]
2\%   &&  1\%     && 0.018  &&  0.003 && 0.019 && 0.019 && 0.018 &&& 0.009 && 0.013 && 0.008 && 0.008 && 0.008 && 0.011\\
      &&  5\%     && 0.077  &&  0.008 && 0.072 && 0.072 && 0.070 &&& 0.005 && 0.010 && 0.005 && 0.005 && 0.005 && 0.007  \\
      &&  10\%    && 0.143  &&  0.018 && 0.134 && 0.133 && 0.131 &&& 0.004 && 0.009 && 0.004 && 0.004 && 0.004 && 0.005    \\
      \multicolumn{14}{c}{} \\ [-3mm]
5\%   &&  1\%     && 0.017  &&  0.002 && 0.018 && 0.018 && 0.017 &&& 0.030 && 0.040 && 0.030 && 0.030 && 0.030 && 0.031 \\
      &&  5\%     && 0.072  &&  0.008 && 0.069 && 0.069 && 0.067 &&& 0.020 && 0.033 && 0.020 && 0.020 && 0.020 && 0.021  \\
      &&  10\%    && 0.133  &&  0.017 && 0.125 && 0.124 && 0.122 &&& 0.015 && 0.030 && 0.016 && 0.016 && 0.016 && 0.017    \\
      \multicolumn{14}{c}{} \\ [-3mm]
10\%  &&  1\%     && 0.016  &&  0.002 && 0.017 && 0.017 && 0.016 &&& 0.074 && 0.089 && 0.075 && 0.075 && 0.075 && 0.074\\
      &&  5\%     && 0.067  &&  0.008 && 0.067 && 0.067 && 0.065 &&& 0.055 && 0.080 && 0.056 && 0.055 && 0.055 && 0.054  \\
      &&  10\%    && 0.122  &&  0.016 && 0.121 && 0.121 && 0.120 &&& 0.044 && 0.074 && 0.045 && 0.045 && 0.045 && 0.045    \\
      \multicolumn{14}{c}{} \\ [-3mm]
20\%   &&  1\%     && 0.014  &&  0.002 && 0.016 && 0.016 && 0.016 &&& 0.166 && 0.187 && 0.169 && 0.169 && 0.169 && 0.171\\
      &&  5\%     && 0.058  &&  0.008 && 0.063 && 0.063 && 0.063 &&& 0.135 && 0.175 && 0.138 && 0.137 && 0.137  &&  0.143 \\
      &&  10\%    && 0.105  &&  0.014 && 0.115 && 0.116 && 0.117 &&& 0.115 && 0.167 && 0.116 && 0.116 && 0.115 &&  0.121   \\
      \multicolumn{14}{c}{} \\ [-3mm]
30\%   &&  1\%     && 0.013  &&  0.002 && 0.015 && 0.016 && 0.016 &&& 0.270 && 0.289 && 0.267 && 0.267 && 0.267 && 0.274\\
      &&  5\%     && 0.051  &&  0.007 && 0.060 && 0.061 && 0.062 &&& 0.238 && 0.278 && 0.231 && 0.230 && 0.229 && 0.235  \\
      &&  10\%    && 0.092  &&  0.013 && 0.110 && 0.113 && 0.116 &&& 0.214 && 0.270 && 0.203 && 0.201 && 0.200 && 0.210    \\
\hline
\multicolumn{14}{l}{\footnotesize{*Type II calculated at optimized Type I error $= \alpha$}}
  \end{tabular}\par}
\label{table:exist_YanZheng} 
\end{table}
\end{landscape}


\subsubsection{Revisiting Yan and Zheng (2017)}
\indent Applying the preferred methods based on Table \ref{table:exist_YanZheng} to the 18,000 anomalies, the fraction of true strategies is found to be 0.000\% (BY) and 0.015\% (Storey, $\theta = 0.8$) under a 5\% significance level, and 0.006\% (BY) and 0.091\% (Storey, $\theta=0.8$) under a 10\% significance level.\footnote{Across different values of $p_0$, BY dominates BH, and Storey ($\theta = 0.8$) dominates the two other Storey methods in achieving the pre-specified significance levels. We therefore focus on BY and Storey ($\theta = 0.8$). The statistics in this paragraph are based on unreported results.
}

Our results suggest that only about 0.1\% (or 18) of the 18,000 anomaly strategies in Yan and Zheng (2017) are classified as ``true" to control the FDR at 10\%. In contrast, Yan and Zheng (2017) claim that ``a large number" of these strategies are true based on the Fama-French (2010) approach. In particular, they suggest that the $90^{th}$ percentile of $t$-statistics is significant, implying that at least 10\% of anomalies, that is, 1,800, generate significant positive returns. They conclude that there exists widespread mispricing.\footnote{In earlier versions of the paper, they find that even the $70^{th}$ percentile is significant (which is confirmed by our analysis), suggesting that the top 30\% of anomalies are true.}

Why are our results so different from those of Yan and Zheng (2017)? First, it is important to realize that the multiple-testing methods that we examine so far (which do not include the Fama-French (2010) approach that Yan and Zheng (2017) employ) have a different objective function than that in Yan and Zheng (2017). In particular, while Yan and Zheng (2017) are interested in finding an unbiased estimate of the fraction of true discoveries, multiple-testing adjustments seek to control the FDR at a certain level, and many true strategies will be missed when the hurdle is high. As a result, it is to be expected that the fraction of true anomalies identified by multiple-testing methods will be somewhat smaller than the true value. However, this is unlikely to fully explain the stark difference between our inference and Yan and Zheng's (2017) estimate, that is, two orders of magnitude. Using our framework, we
show that their application of the Fama-French (2010) approach is problematic.

More specifically, we show that the Fama-French (2010) approach is not designed to estimate the fraction of true strategies. As we explore in Section II.C, the Fama-French (2010) approach focuses only on whether the entire population of strategies has a zero mean return. When the null hypothesis (i.e., the entire population has a zero mean) is rejected, all we know is that some strategies are true --- we cannot tell how many are true. While we focus on the misapplication of the Fama-French (2010) approach in estimating the fraction of true strategies here, we defer the analysis of its test power to Section II.C.

To apply our framework, we assume that a fraction $p_0$ of strategies are true for their anomalies. Next, we define the test statistic that describes the inference approach that is implicitly used in Yan and Zheng (2017). Let $Frac_d$ be the $(100-d)^{th}$ percentile of the cross-section of $t$-statistics and let $p(Frac_d)$ be the $p$-value associated with $Frac_d$ based on the bootstrapping approach in Fama and French (2010). Define $Frac$ as the maximum $D$ in $(0,0.4)$ such that $p(Frac_d) \leq 0.05$ is true for all $d\in(0, D)$.\footnote{More formally, $Frac = \max_{D\in(0,0.4)} \{\max\{p(Frac_d)\}_{d\leq D} \leq 0.05\}$.} In other words, $D$ is the maximum fraction such that all percentiles above $100(1-D)$ are rejected at the 5\% level based on the Fama-French (2010) approach. We set the upper bound at 0.4 because it is unlikely that more than 40\% of their anomalies generate significantly positive returns given the distribution of $t$-statistics shown in Figure \ref{fig:ROC}. Changing the upper bound to alternative values does not qualitatively affect our results.

\begin{table}[ht]
\centering 
  \footnotesize\addtolength{\tabcolsep}{-4pt}
  \caption{\textbf{Diagnosing Yan and Zheng (2017): Fraction of Anomalies Identified}}
\captionsetup{width=13.0cm}
\caption*{\textmd{\small{This table presents simulated fraction of anomalies identified by following the Fama-French (2010) approach in Yan and Zheng (2017).} For each fraction $p_0$ of strategies that are believed to be true, we follow our method in Section I and set $I = 100$ (for each $i$, we bootstrap to obtain the ranking of strategies and set the top $p_0$ as true) and $J = 100$ (conditional on $i$, for each $j$, we bootstrap the time periods) to run 10,000 ($=100\times 1,00$) bootstrapped simulations. For each bootstrap sample, we perform the Fama-French (2010) test and find $Frac$, the maximum fraction $D$ such that all percentiles above $100(1-D)$ (with 1\% increment) are rejected at the 5\% level. We set the upper bound of $D$ at 40\%.}}
\small{
\begin{tabular}{ccccccccccccc} 
\hline\hline
      &&& \multicolumn{9}{c}{Summary Statistics for $Frac$} \\
      \cline{4-13}
$p_0$ &&& Mean (\%) && Stdev.(\%) &&& Prob($Frac \geq 5\%$) &&  Prob($Frac \geq 10\%$) \\
\hline
\multicolumn{13}{c}{} \\ [-3mm]
0.5\%      &&&  2.06  && 7.31 &&& 0.18   && 0.15 \\
\multicolumn{13}{c}{} \\ [-4mm]
1.0\%      &&&  4.04 && 9.04 &&&  0.23  && 0.21 \\
\multicolumn{13}{c}{} \\ [-4mm]
2.0\%      &&& 7.30 && 10.27 &&&  0.44 && 0.31 \\
\multicolumn{13}{c}{} \\ [-4mm]
5.0\%      &&& 20.19 && 13.36 &&&  0.99 && 0.83\\
\multicolumn{13}{c}{} \\ [-4mm]
10.0\%     &&& 35.68 && 7.55 &&& 1.00 && 1.00\\
\multicolumn{13}{c}{} \\ [-4mm]
15.0\%     &&& 39.62 && 1.76 &&& 1.00 &&1.00  \\
\hline
  \end{tabular}\par}
\label{table:diagnose_YanZheng} 
\end{table}

Table \ref{table:diagnose_YanZheng} reports summary statistics on $Frac$ for different levels of $p_0$. The simulated means of $Frac$ all exceed the assumed levels of $p_0$, which suggests that the approach taken by Yan and Zheng (2017) is biased upward in estimating the fraction of true anomalies. Focusing on the last column (i.e., the probability of $Frac$ exceeding 10\%), when $p_0$ is only 0.5\%, the probability that the Fama-French (2010) test statistic would declare all top 10\% of anomalies true is 15\%. If $p_0$ were 10\% (as claimed by Yan and Zheng (2017)), the Fama-French (2010) test statistic would frequently declare more than 30\% of anomalies true (as seen from a mean statistic of 35.68, which is close to the 40\% upper bound).

The intuition for the results above is straightforward. Suppose $p_0 = 10\%$. Due to sampling uncertainty, not all 10\% of true strategies will be ranked above the $90^{th}$ percentile. Given the large number of tests for more than 18,000 anomalies, strategies with a population mean of zero may be lucky and effectively crowd out true strategies by being ranked higher than the $90^{th}$-percentile $t$-statistic for a given sample. As such, the 10\% of strategies that are true affect not only the $90^{th}$ percentile of $t$-statistics and beyond, but also possibly lower percentiles. The Fama-French (2010) approach may therefore detect significant deviations from the null hypothesis for a lower percentile, thereby overestimating the fraction of true strategies.

To summarize, Yan and Zheng (2017) misapplies the Fama-French (2010) approach to reach the conclusion that a large number (i.e., more than 1,800) of anomalies are true. Using multiple-testing methods that are most powerful in detecting true anomalies (while controlling the FDR at the 10\% level), we find that a very small number (i.e., 18) are significant. While additional investigation of the economic underpinnings of these significant anomalies may further narrow the list of true anomalies, our results cast doubt on the possibility of discovering true anomalies through a pure data-mining exercise, such as that carried out in Yan and Zheng (2017).

%

\subsection{A Simulation Study}
\indent We now perform a simulation study to evaluate the performance of our proposed method. Similar to our application when $p_0$ is unknown, we focus on the ability of our method to correctly rank the performance of existing methods. We are particularly interested in how the cross-sectional alpha (or mean return) distribution among true strategies affects our results. We therefore examine a variety of cross-sectional distributions, which we denote by $F$.

We also seek to highlight the Type I versus Type II error trade-off across existing multiple-testing methods. However, since hypothesis testing is essentially a constrained optimization problem, where we face a Type I error constraint (i.e., Type I error rate is no greater than the pre-specified significance level) while seeking to maximize test power (i.e., one minus the Type II error rate), a tighter constraint (e.g., the pre-specified significance level is exactly met) leads to a lower Type II error rate. It is therefore sufficient to show how our method helps select the best-performing method in terms of meeting the Type I error rate constraint.\footnote{Conditional on the Type I error rate being less than the desired significance level, the closer the Type I error rate is to the desired level, the smaller the Type II error rate will be. Therefore, given the lack of a specification for the relative weights of Type I and Type II error rates, we do not report results on the Type II error rates. In addition, in contrast to the Type I error rate, we do not have a good benchmark for thinking about the magnitude of the Type II error rate. For instance, if 10\% is the desired significance level for the Type I error rate, an actual rate of 1\% would be considered too low. For Type II errors, suppose the true error rate is 0.5\% (which is roughly consistent with the magnitude of the Type II error rate we observe in the data given our definition) --- would an actual rate of 0.1\% be considered very different? While its distance from the true error rate is small in absolute terms, it is high in percentage terms.
}

We conduct our simulation study using the CAPIQ data.\footnote{We focus on the CAPIQ data to save computational time. Our simulation exercise is computationally intensive. Each simulation run takes on average one day on a single core. The computing platform that we have access to allows us to have 400 cores running at the same time. It therefore takes us one day to complete a specification (i.e., F) with 400 independent simulation runs. It would take us much longer for the data with 18,000 anomalies.
} To preserve the cross-sectional dependence in the data, we simulate strategy returns based on the actual CAPIQ data. We set the fraction of true strategies to 10\% throughout our analysis. However, our specification of $p_0$ does not have to equal 10\% (i.e., does not need to be correctly specified). We also study alternative values of $p_0$ to examine our model's performance when $p_0$ is potentially misspecified.

We explore several specifications for $F$. They all take the form of a Gamma distribution with mean $\mu_0$ and standard deviation $\sigma_0$.\footnote{The probability density function for a Gamma-distributed variable $X$ with mean $\mu_0$ and standard deviation $\sigma_0$ is $f(x) = \frac{1}{\Gamma(k)\theta^k}x^{k-1}e^{-\frac{x}{\theta}}$, where the shape parameter is $k = \frac{\mu_0^2}{\sigma_0^2}$, the scale parameter is $\theta = \frac{\sigma_0^2}{\mu_0}$, and $x>0$.} We entertain three values for $\mu_0$ ($\mu_0 = 2.5\%$, $5\%$, and $10\%$) and three values for $\sigma_0$ ($\sigma_0 = 0$, $2.5\%$, and $5\%$).\footnote{In the data, the mean return for the top 10\% of strategies ranked in terms of the $t$-statistic (which is equivalent to the Sharpe ratio in our context) is 6.3\%.} (For example, $(\mu_0 = 5.0\%, \sigma_0 = 0)$ denotes a constant (i.e., a point mass distribution) at $\mu_0 = 5.0\%$.) In total, we study nine specifications for $F$.

There are several advantages to modeling the cross-sectional distribution $F$ with a Gamma distribution. First, we focus on one-sided tests in this section, so $F$ should have positive support. Second, moments are available analytically, so we can easily change the mean (variance) while holding the variance (mean) constant to study the comparative statics. Finally, a Gamma distribution has higher-moment characteristics (e.g., skewness and excess kurtosis) that allow it to capture salient features of $F$.\footnote{For our parameterization of the Gamma distribution (mean $\mu_0$ and standard deviation $\sigma_0$), skewness is $\frac{2\sigma_0}{\mu_0}$ and excess kurtosis is $\frac{6\sigma_0^2}{\mu_0^2}$.}

Our simulation exercise is conducted as follows. For the CAPIQ data, we first randomly select 10\% of strategies and inject mean returns that are generated by the distribution $F$. For the remaining 90\% of strategies, we set their mean returns to the null hypothesis (i.e., zero). Let $D_m$ denote the final data, where $m$ denotes the $m$-th iteration. In the next step of the analysis, $D_m$ can be thought of as the population of strategy returns.

We next perturb $D_m$ to generate the in-sample data. In particular, we bootstrap the time periods once while keeping the cross section intact to generate the in-sample data $D_{m,k}$, where the subscript $k$
represents the round of iteration conditional on $m$. Note that each in-sample data matrix $D_{m,k}$ is generated conditional on the same population $D_m$.

For each $D_{m,k}$, we follow a given multiple-testing method (e.g., BH) for a pre-specified significance level $\delta$ to generate the test outcomes. Since we know the true identities of strategies based on the step at the beginning of the simulation when we create $D_m$, we are able to calculate the true realized error rates, which we denote by $FDR^a_{m,k}$ (``\emph{a}'' stands for ``actual''). This actual realized error will vary across the different multiple-testing methods as well as the different nominal error rates. Implementing our method for $D_{m,k}$, for which we do not know the true strategies, we get the estimated error rates $FDR^e_{m,k}$ (``\emph{e}'' stands for ``estimated'').

We report three statistics in our results: the nominal error rate ($\delta$, which is also the pre-specified significance level), the actual error rate (\textit{Actual}), and our estimated error rate (\textit{Est.}). The actual error rate and our estimated error rate are given by
\begin{eqnarray*}
\text{\emph{Actual}} &=& \frac{1}{MK}\sum_{m=1}^M \sum_{k=1}^K FDR^a_{m,k}, \\
\text{\emph{Est.}} &=& \frac{1}{MK}\sum_{m=1}^M \sum_{k=1}^K FDR^e_{m,k}. \\
\end{eqnarray*}

For a multiple-testing method, the pre-specified significance level (i.e., $\delta$) can be viewed as the stated significance level, which can be different from the true error rate. The difference between $\delta$ and \textit{Actual} therefore captures the bias in error rate committed by the method. This bias can be revealed through our approach if \textit{Est.} is reasonably close to \textit{Actual}, in particular, if \textit{Est.} is closer to \textit{Actual} than $\delta$. If our approach provides good approximations to the true error rates across all specifications, then one can use our method to rank existing methods in terms of their performance with respect to the Type I error rate.

Table \ref{table:simu_constant} reports the results when the cross-sectional distribution $F$ is assumed to be a constant. Overall, our method presents a substantial improvement over the usual approach that takes the stated significance level as given. Our results show that \textit{Est.} is often much closer to \textit{Actual} than $\delta$.

The performance of our method in approximating the true error rate improves when the mean parameter $\mu_0$ is higher. Note that a higher $\mu_0$ (and hence a higher signal-to-noise ratio in the data) does not necessarily lead to better performance for existing multiple-testing methods. For example, at the 10\% significance level, while BH improves when $\mu_0$ increases from 2.5\% to 10\% (i.e., as the true Type I error rate gets closer to 10\%), the performance of Storey ($\theta=0.8$) declines as its true Type I error rate increases from 10.29\% to 12.09\%. Our approach performs better when $\mu_0$ is higher (a higher signal-to-noise ratio) because it is easier for our first-stage bootstrap to correctly identify the true outperforming strategies, leading to a more accurate calculation of error rates in the second-stage bootstrap. The two reasons above explain the larger improvement of our method over the usual approach when the signal-to-noise ratio is higher in the data.

Our estimation framework also appears to be robust to potential misspecification in $p_0$. There is some evidence that a correctly specified $p_0$ (i.e., $p_0 = 10\%$) leads to better performance on average across all scenarios (i.e., across different methods, significance levels, and specifications of $\mu_0$). However, for a particular specification, this may not be the case. For example, for BH and at the 5\% level, given a true error rate of 0.0472, a misspecified $p_0$ of 5\% seems to perform somewhat better ($Est. = 0.0465$) than $p_0 = 10\%$ ($Est. = 0.0459$). Nonetheless, the performance of our method in approximating the true error rates is consistent across alternative specifications of $p_0$.


Tables IA.I and IA.II in Section I of the Internet Appendix report the results under alternative specifications of the cross-sectional distribution $F$.\footnote{The Internet Appendix is in the online version of the article on the Journal of Finance website.} Compared to Table \ref{table:simu_constant}, larger dispersion in the mean return distribution (while keeping the mean constant) seems to generate a higher Type I error rate for BH and BY when $\mu_0$ is low (i.e., $\mu_0 = 2.5\%$). This can be explained as follows. True strategies with a low signal-to-noise ratio (e.g., strategies that have a mean return below 2.5\%) generate lower $t$-statistics in comparison with those in Table \ref{table:simu_constant}. This leads BH and BY to lower the $t$-statistic cutoff to maintain the pre-specified significance level, which makes it easier for null strategies to overcome this threshold, and leads in turn to a higher Type I error rate. For these cases, our approach performs even better than in Table \ref{table:simu_constant} in estimating the true error rates. It performs similarly to Table \ref{table:simu_constant} for other specifications.

It is worth noting that while the performance of existing multiple-testing methods varies across different specifications, our approach consistently compares favorably with most methods. For example, in Table \ref{table:simu_constant} under $\mu_0 = 2.5\%$, Storey ($\theta = 0.8$) works well under a significance level of 10\% as its true Type I error rate (i.e., 10.29\%) is close to the desired level. As a result, the stated significance level of $\alpha = 10\%$ is a reasonably good estimate of the true error rate. However, the performance of Storey ($\theta = 0.8$) deteriorates substantially when $\mu_0$ is raised to $5\%$ or $10\%$, leading to large estimation errors when using the stated significance level to approximate the true error rates. In contrast, our approach performs well across all specifications of $\mu_0$, making it the robust choice when assessing the performance of Storey ($\theta = 0.8$).

We also note that our simulation analysis focuses on the CAPIQ data, which are the actual data we use in our empirical work in this paper. For researchers interested in applying our approach, we also recommend using our simulation design to evaluate the performance of our method for the specific data under consideration.\footnote{Given the nature of our approach, we particularly caution against applications of our approach where a very small $p_0$ is chosen and the signal-to-noise ratio in the data is low. In such cases, estimation error is high so the top $100p_0$\% of strategies may be outliers whose performance is driven by estimation errors rather than information in the data. As such, the in-sample estimates of performance for these strategies may be very different from their population values, leading to biased estimates of Type I and Type II errors. We recommend a user-specific simulation study that resembles ours to determine the applicability of our approach for these cases.
}

\begin{landscape}
\begin{table}[ht]
\centering 
  \footnotesize\addtolength{\tabcolsep}{-4pt}
  \caption{\textbf{A Simulation Study on CAPIQ: Mean Return Distribution for True Strategies Is A Constant }  }
\captionsetup{width=23.0cm}
\caption*{\textmd{\small{This table presents simulated Type I error rates for CAPIQ when the mean return distribution for true strategies ($F$) is a constant.} The simulation study proceeds as follows. We fix the fraction of true strategies at 10\%. We first randomly identify 10\% of strategies as true and assign mean returns to them according to $F$. Mean returns are set to zero for the remaining 90\% of strategies. Let $D_m$ denote the final data ($m=1,2,\ldots, M = 400$). Conditional on $D_m$, we bootstrap the time periods to generate the perturbed in-sample data $D_{m,k}$ ($k=1,2,\ldots, K = 100$). For each $D_{m,k}$ and for a given multiple-testing method at a pre-specified significance level $\delta$, we calculate the true realized error rate ($FDR^a_{m,k}$). Implementing our approach (with a prior specification of $p_0$), we obtain the estimated error rate ($FDR^e_{m,k}$). We report the mean true Type I error rate (\textit{Actual}), defined as $\frac{1}{MK}\sum_{m=1}^M \sum_{k=1}^K FDR^a_{m,k}$, and mean error rate for our estimator (\textit{Est.}), defined as $\frac{1}{MK}\sum_{m=1}^M \sum_{k=1}^K FDR^e_{m,k}$. $\delta$ denotes the nominal error rate. The distribution $F$ is a point mass at $\mu_0$, where $\mu_0$ takes the value of 2.5\%, 5.0\% or 10\%. * indicates better model performance, that is, \textit{Est.} is closer to \textit{Actual} compared to $\delta$. }}
\hspace*{-0.5cm}
\small{
\begin{tabular}{ccccccccccccccccccccccccccccc} 
\hline\hline
    \multicolumn{29}{c}{} \\ [-3mm]
      && &&\multicolumn{7}{c}{$\mu_0 =2.5\%$} &&& \multicolumn{7}{c}{$\mu_0 =5.0\%$}  &&& \multicolumn{7}{c}{$\mu_0 =10\%$}\\
        \multicolumn{29}{c}{} \\ [-3mm]
      \cline{5-11} \cline{14-20} \cline{23-29}
     \multicolumn{29}{c}{} \\ [-3mm]
Method  && $\delta$ && \emph{Actual } && \multicolumn{5}{c}{\emph{Est.}} &&& \emph{Actual}  && \multicolumn{5}{c}{\emph{Est.}} &&& \emph{Actual} && \multicolumn{5}{c}{\emph{Est.}} \\
  \multicolumn{29}{c}{} \\ [-3mm]
\cline{7-11} \cline{16-20} \cline{25-29}
 \multicolumn{29}{c}{} \\ [-3mm]
        &&          &&        && ${\scriptstyle p_0 = 5\%}$ && ${\scriptstyle p_0 = 10\%}$ && ${\scriptstyle p_0 = 15\%}$ &&&         &&    ${\scriptstyle p_0=5\%}$&& ${\scriptstyle p_0 = 10\%}$&&  ${\scriptstyle p_0 = 15\%}$  &&&      &&${\scriptstyle p_0=5\%}$ &&${\scriptstyle p_0=10\%}$ &&${\scriptstyle p_0=15\%}$ \\
        \hline
         \multicolumn{29}{c}{} \\ [-2mm]
BH      &&   1\%  && 0.0075     && 0.0102&& 0.0097$^*$&& 0.0091$^*$      &&& 0.0104                   && 0.0111$^*$ && 0.0105$^*$ &&  0.0100$^*$ &&&  0.0108                  && 0.0116$^*$ && 0.0109$^*$&& 0.0095$^*$    \\
        &&   5\%  &&  0.0375                      && 0.0425$^*$&&0.0410$^*$&&0.0390$^*$       &&& 0.0472                && 0.0465$^*$ && 0.0459$^*$&& 0.0445$^*$  &&& 0.0461     && 0.0480$^*$&& 0.0463$^*$ && 0.0438$^*$   \\
        &&   10\% &&  0.0718                      && 0.0794$^*$&& 0.0771$^*$&& 0.0737$^*$      &&&  0.0876         && 0.0872$^*$ && 0.0845$^*$&& 0.0804$^*$  &&& 0.0875      && 0.0897$^*$&& 0.0873$^*$&& 0.0827$^*$   \\
       \multicolumn{29}{c}{} \\ [-2mm]
BY      &&   1\%  &&    0.0016             && 0.0025$^*$&& 0.0024$^*$&& 0.0022$^*$     & && 0.0020            && 0.0023$^*$ && 0.0021$^*$ && 0.0022$^*$     & && 0.0019              && 0.0023$^*$ && 0.0021$^*$&& 0.0020  \\
        &&   5\%  &&    0.0059                  && 0.0089$^*$&&0.0085$^*$ && 0.0080$^*$     &&& 0.0080                && 0.0081$^*$&& 0.0080$^*$&& 0.0075$^*$     &&&  0.0083     && 0.0089$^*$&& 0.0083$^*$ && 0.0079$^*$   \\
        &&   10\% &&   0.0108                    && 0.0157$^*$&& 0.0150$^*$&& 0.0142$^*$      &&&  0.0155                && 0.0155$^*$ && 0.0150$^*$&& 0.0142$^*$  &&& 0.0154    && 0.0163$^*$&& 0.0154$^*$ && 0.0146$^*$   \\
        \multicolumn{29}{c}{} \\ [-2mm]
Storey   &&   1\%  &&  0.0127      && 0.0138$^*$&& 0.0140$^*$&& 0.0141$^*$     &&& 0.0137   && 0.0153$^*$&& 0.0150$^*$ && 0.0147$^*$     &&&   0.0151        && 0.0153$^*$ && 0.0141$^*$ && 0.0117$^*$   \\
 ($\theta = 0.4$)       &&   5\%  &&    0.0525          && 0.0548$^*$ && 0.0545$^*$ && 0.0544$^*$     &&& 0.0575                && 0.0598$^*$&& 0.0596$^*$&& 0.0584$^*$     & &&   0.0615           && 0.0614$^*$ && 0.0608$^*$&& 0.0599   \\
        &&   10\% &&    0.0950         && 0.0988$^*$&& 0.0990$^*$&& 0.0984$^*$      &&& 0.1049                   && 0.1080$^*$&& 0.1083$^*$ && 0.1064$^*$     &&&   0.1129   && 0.1125$^*$&& 0.1110$^*$&& 0.1095$^*$  \\
        \multicolumn{29}{c}{} \\ [-2mm]
Storey  &&   1\%  &&   0.0124       && 0.0143$^*$&& 0.0141$^*$&& 0.0138$^*$    & &&  0.0135                  && 0.0147$^*$&& 0.0146$^*$ && 0.0143$^*$   & &&  0.0147      && 0.0148$^*$&& 0.0146$^*$&& 0.0144 $^*$   \\
($\theta = 0.6$)         &&   5\%  &&    0.0524          && 0.0542$^*$&& 0.0544$^*$&& 0.0547$^*$      &&& 0.0574                 && 0.0592$^*$&& 0.0595$^*$&& 0.0588$^*$      &&&   0.0616             && 0.0598$^*$&& 0.0607$^*$ && 0.0601   \\
        &&   10\% &&   0.0981               && 0.1002&& 0.1004&& 0.1007       &&&  0.1073                   && 0.1085$^*$&& 0.1099$^*$&& 0.1091$^*$      &&&   0.1147             && 0.1100$^*$&& 0.1129$^*$ && 0.1121$^*$  \\
        \multicolumn{29}{c}{} \\ [-2mm]
Storey   &&   1\%  &&  0.0114          && 0.0121$^*$&& 0.0121$^*$&& 0.0120$^*$     & &&  0.0127      && 0.0135$^*$&& 0.0136$^*$ && 0.0135$^*$      &&&  0.0136                && 0.0135$^*$&& 0.0136$^*$ && 0.0136$^*$   \\
($\theta = 0.8$)        &&   5\%  &&   0.0532                 && 0.0537$^*$&& 0.0549$^*$&& 0.0552$^*$      &&& 0.0576                  && 0.0581$^*$&& 0.0594$^*$&& 0.0599$^*$     &&&   0.0618               && 0.0588$^*$&& 0.0609$^*$&&0.0615$^*$    \\
        &&   10\% &&    0.1029           && 0.1022$^*$&& 0.1050$^*$ && 0.1058$^*$     & && 0.1137                   && 0.1113$^*$&& 0.1157$^*$&& 0.1174$^*$    & &&   0.1209    && 0.1132$^*$&& 0.1193$^*$&& 0.1212$^*$    \\
\hline
  \end{tabular}\par}
\label{table:simu_constant} 
\end{table}
\end{landscape}

\subsection{Performance Evaluation}

\subsubsection{Data Description: Mutual Funds}
\indent Our mutual fund data are obtained from the Center for Research in Security Prices (CRSP) Mutual Fund database. Since our goal is to take another look at Fama and French (2010), we apply similar screening procedures to Fama and French (2010) to obtain our data set. In particular, our data start from January 1984 to mitigate omission bias (Elton, Gruber, and Blake (2001)) and end in December 2006 to be consistent with Fama and French (2010). We also limit tests to funds with assets under management (AUM) of at least five million dollars in 2006 terms --- once a fund passes this size threshold, its subsequent returns are included in our tests. We also require that a fund have at least eight monthly return observations to be included in our tests.\footnote{Notice that this requirement applies to funds in both the actual sample and the bootstrapped samples. We examine alternative cutoffs in Section II.C.4.} Finally, following Fama and French (2010), we restrict our sample to funds that appear on CRSP at least five years before the end of the sample period (i.e., 2006) to avoid funds that have a short return history. We use the four-factor model of Fama and French (1993) and Carhart (1997) as our benchmark factor model applied to net mutual fund returns.\footnote{Both the Fama-French (1993) factors and the momentum factor are obtained from Ken French's online data library. }

Our mutual fund sample closely follows that used by Fama and French (2010). For example, 3,156 funds in Fama and French's (2010) sample that have initial AUM exceeding five million 2006 dollars. In our sample, we have 3,030 such funds. Summary statistics on fund returns are reported in Section II of the Internet Appendix. For our main results, we focus on the full sample of funds, that is, all funds that have initial AUM exceeding five million 2006 dollars. We examine alternative samples (i.e., AUM $=$ \$250 million and \$1 billion) in Section II of the Internet Appendix.

\begin{figure}[H]
\vspace*{-2cm}
\centering
\includegraphics[trim=4.1cm 4.5cm 1cm 3.0cm, clip= true,width= 1.2\textwidth, height = 0.6\textwidth]{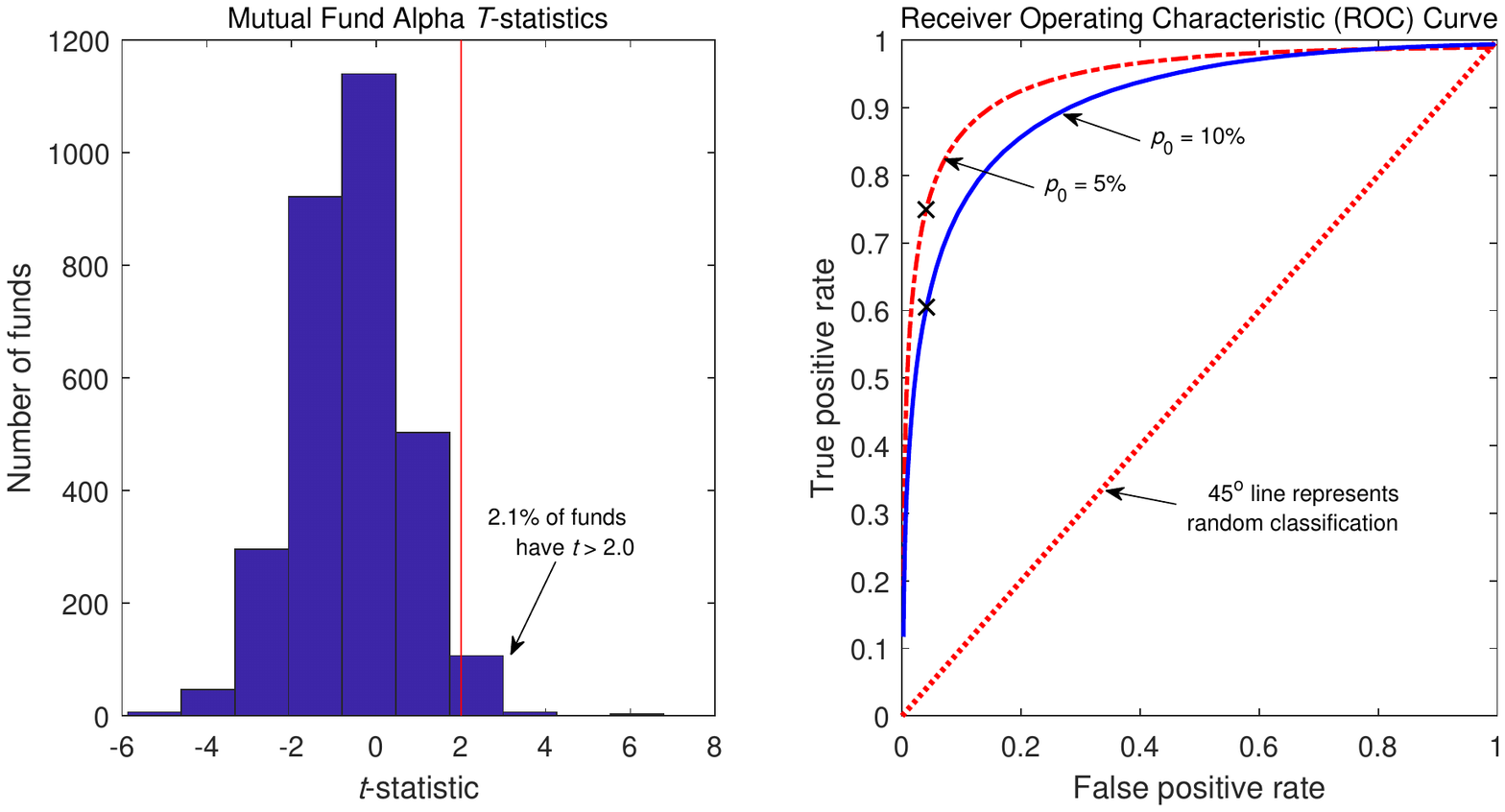}
\vspace*{-1.0cm}
\caption{\textbf{Mutual fund performance.} \textmd{\footnotesize{This figure shows the $t$-statistic distribution and the receiver operating characteristic (ROC) curve for mutual fund alphas.} For each fund in our data (1984 to 2006), we calculate the $t$-statistic of its alpha corresponding to the Fama-French-Carhart four-factor model. The left figure plots the $t$-statistic distribution. The right figure plots the ROC curves corresponding to $p_0 = 5\%$ and $p_0 = 10\%$. The crosses on the ROC curves mark the points that correspond to a $t$-statistic of 2.0.
}}
\label{fig:dist_MF}
\end{figure}

Figure \ref{fig:dist_MF} shows the $t$-statistic distribution as well as the ROC curve for mutual fund alphas. Compared to the $t$-statistic distribution for the CAPIQ data in Figure \ref{fig:ROC}, the fraction of funds with large and positive $t$-statistics is much smaller. This difference is reflected in the ROC curve, where under the same $p_0$ (i.e., $p_0 = 10\%$), the ROC curve for the CAPIQ data is pushed more towards the northwest direction than that for mutual funds, indicating a higher true positive rate for the same level of false positive rate. The $t$-statistic distribution for mutual funds is also skewed to the left, which is consistent with previous findings that there is more evidence for extreme underperformers than for extreme outperformers. For the purpose of our application, we focus on the right tail of the distribution.

\subsubsection{Luck vs. Skill for Mutual Fund Managers}
\indent In contrast to our previous applications, we study a different question for the mutual fund data. In particular, we reexamine the question of whether there exist any outperforming funds. We focus on the joint test approach used in Fama and French (2010), which treats the mutual fund population as a whole and tests whether the entire mutual fund population has zero alpha (the null hypothesis) versus at least one fund has a positive alpha. Note that the goal of the joint test is different from the goal of multiple testing studied above, which is to identify the fraction of outperforming funds. We deliberately choose an application that is different from multiple tests to highlight the generality of our framework. We also use our framework to evaluate the Fama and French (2010) approach, which is an important and popular method in the literature.\footnote{For recent papers that apply the Fama and French (2010) approach or its predecessor, KTWW (2006), see Chen and Liang (2007), Jiang, Yao, and Yu (2007), Ayadi and Kryzanowski (2011), D'Agostino, McQuinn, and Whelan (2012), Cao et al. (2013), Hau and Lai (2013), Blake et al. (2013), Busse, Goyal, and Wahal (2014), Harvey and Liu (2017), and Yan and Zheng (2017). } Fama and French's (2010) joint test suggests that very few (if any) funds exhibit skill on a net return basis.



More specifically, we use our double-bootstrap approach to evaluate the Type I and Type II error rates for the single-bootstrap approach proposed by Fama and French (2010). As we explain in Section I.D, the Type I and Type II error rates in this context correspond to the probability of rejecting the null hypothesis of zero performance across all funds when this null is true (Type I error rate) and the probability of not rejecting this null when some funds have the ability to generate a positive alpha (Type II error rate). By varying $p_0$ (the fraction of funds that generate a positive alpha), we calculate these two error rates for the Fama and French (2010) approach.


While $p_0$ captures the prior belief about the fraction of skilled managers, it does not tell us the magnitude of the average alpha across these managers. We therefore calculate two additional statistics that help gauge the economic significance of the alphas generated by outperforming funds. In particular, \textit{Avg. alpha} and \textit{Avg. t-stat of alpha} are the average (across simulations) alpha and the $t$-statistic of alpha for the fraction $p_0$ of funds that are assumed to be outperforming.\footnote{We calculate \textit{Avg. alpha} and \textit{Avg. t-stat of alpha} as follows. Following the four-step procedure of our method presented in Section I.B, for each bootstrapped sample as generated by Step I, we find the top $p_0$ of funds and calculate the median alpha and the $t$-statistic of alpha. We then take the average across $J$ bootstrapped samples (Step III) to calculate \textit{Avg. alpha} and \textit{Avg. t-stat of alpha}.}\textsuperscript{,}\footnote{Note that due to the undersampling problem of the Fama and French (2010) approach that we identify below, the reported \textit{Avg. alpha} and \textit{Avg. t-stat of alpha} are somewhat higher than the average alpha and average $t$-statistic of alpha for the actual funds in the bootstrapped sample. However, had smaller alphas been injected into funds in our first-stage bootstrap, the test power for the Fama and French (2010) approach would be even lower. Therefore, one can interpret our results as providing upper bounds on test power for a given average level of alpha under the alternatives. }

Table \ref{table:sum_FF} presents the simulated Type I and Type II error rates for Fama and French (2010), as well as the average alpha and the $t$-statistic of alpha for outperforming funds under different values of $p_0$. We examine four of the extreme percentile statistics (i.e., $90^{th}$, $95^{th}$, $98^{th}$, and $99^{th}$ percentiles) used in Fama and French (2010), as well as two additional extreme percentiles (i.e., $99.9^{th}$ and $99.5^{th}$) and the max statistic. Below we highlight the difference between the test statistics used in Fama and French (2010) and the additional test statistics we consider. The use of extreme percentiles in this context is driven by the general idea that extreme percentiles are more sensitive to the existence of outperforming funds and therefore are potentially more powerful in detecting outperformance. For example, if 1\% of funds are outperforming, we should expect to see a statistically significant $99^{th}$ percentile (as well as $99.9^{th}$ and $99.5^{th}$ percentiles) when tested against its distribution under the null hypothesis. In contrast, the other percentiles should be insignificant or at least less significant. In Table \ref{table:sum_FF}, a given column (which corresponds to a certain test statistic) reports the Type I error rates (Panel A) and the Type II error rates (Panel B) associated with different significance levels. If the test statistic performs well, the Type I error rates should be lower than the pre-specified significance levels and the Type II error rates should be close to zero.

\begin{table}[ht]
\centering 
  \footnotesize\addtolength{\tabcolsep}{-4pt}
  \caption{\textbf{Simulated Error Rates for Fama and French (2010)}  }
\captionsetup{width=16.0cm}
\caption*{\textmd{\small{This table presents simulated Type I and Type II error rates for the Fama and French (2010) approach.} For each fraction $p_0$ of strategies that are believed to be true, we follow our method in Section I.D and perturb the adjusted data $M = 1,000$ times to calculate the empirical Type I and Type II error rates. We consider six percentiles as well as the maximum. \textit{Avg. alpha} and \textit{Avg. t-stat of alpha} report the average (across simulations) of the mean alpha and alpha $t$-statistic for the fraction $p_0$ of funds that are assumed to be outperforming. All funds with initial AUM exceeding \$5 million are included, resulting in 3,030 funds over the 1984 to 2006 period. Panel A reports the Type I error rate, that is, the probability of falsely declaring that some managers have skill when no fund is outperforming. Panel B reports the Type II error rate, that is, the probability of falsely declaring that no manager is skilled when some funds are outperforming. }}
\footnotesize{
\begin{tabular}{ccccccccccccccccccccccc} 
\hline\hline
      &&      &&             &&          && \multicolumn{13}{c}{Test Statistics (Percentiles)} \\
      \cline{9-23}
$p_0$ && \emph{Avg.} && \emph{Avg. t-stat} && $\alpha$ && Max && 99.9\% && 99.5\% && 99\% && 98\% && 95\% && 90\%  \\
(frac. of true) && \emph{alpha}   && \emph{of alpha} && (sig. level) && && && && && && && && \\
\hline
\multicolumn{23}{c}{} \\ [-3mm]
\multicolumn{23}{c}{Panel A: Type I Error Rate} \\
\hline
\multicolumn{23}{c}{} \\ [-3mm]
0  && 0 && 0  &&  1\%    &&0.001 && 0.000 && 0.003  &&  0.000      &&   0.002     &&   0.009      &&  0.014           \\
   &&&&   &&  5\%       && 0.017 && 0.003 && 0.004    && 0.005        && 0.016    && 0.029        &&   0.040           \\
   &&&&   &&  10\%      && 0.033 && 0.009 && 0.004   &&  0.007       &&  0.043      &&  0.071       &&    0.079       \\
      \hline
\multicolumn{23}{c}{} \\ [-3mm]
\multicolumn{23}{c}{Panel B: Type II Error Rate} \\
\hline
\multicolumn{23}{c}{} \\ [-3mm]
0.5\%  && 14.60 && 3.97  &&  1\%   && 0.998 && 1.000 && 0.997 &&  0.999   && 0.997       &&  0.988       &&  0.983          \\
     &&&& &&  5\%                 && 0.977 && 0.997 && 0.997 && 0.994   &&   0.978   &&  0.965        &&   0.960          \\
     &&&& &&  10\%         &&   0.953   && 0.989 && 0.993 &&  0.980   &&  0.946       &&  0.920       &&  0.916        \\
\multicolumn{23}{c}{} \\ [-3mm]
1\%  && 13.22 && 3.68  &&  1\%  && 0.997 && 1.000 && 0.999 &&  1.000    &&   0.996     &&   0.985      &&  0.981           \\
     &&&&   &&  5\%            &&  0.970   && 0.996 && 0.997 &&  0.991  && 0.971    && 0.959        &&   0.957           \\
     &&&& &&  10\%          &&  0.942   &&  0.986 && 0.990 &&  0.966  &&  0.932     &&  0.912       &&    0.910        \\
\multicolumn{23}{c}{} \\ [-3mm]
2\%  && 10.66 && 3.30  &&  1\%      && 1.000 && 0.999 && 0.998 &&  0.982        && 0.984      &&  0.983       &&    0.981        \\
     &&&& &&  5\%           &&  0.976     && 0.994 && 0.993 &&  0.869 && 0.937    &&  0.938       &&   0.948         \\
     &&&& &&  10\%         &&   0.957    && 0.986 && 0.983 && 0.741 &&  0.852      &&  0.880      && 0.898        \\
\multicolumn{23}{c}{} \\ [-3mm]
3\%  && 9.75 && 3.02  &&  1\%     && 0.997 && 0.997&& 0.999  &&  0.867        &&   0.914     &&   0.976      &&  0.975           \\
     && && &&   5\%           && 0.966   && 0.991 && 0.992 &&   0.478  && 0.677    && 0.896        &&   0.925           \\
     &&&& &&  10\%          && 0.940   && 0.981 && 0.965 &&  0.277  &&  0.506      &&  0.795       && 0.849        \\
\multicolumn{23}{c}{} \\ [-3mm]
5\% && 8.13 && 2.63   &&  1\%    && 0.998 && 1.000 && 0.998 &&  0.802     &&  0.710     &&  0.961       && 0.969           \\
    &&&&  &&  5\%        &&  0.974  && 0.995 && 0.992 &&  0.363 && 0.336    && 0.815        &&  0.898          \\
    &&&&  &&  10\%        &&  0.947 && 0.983 && 0.957 &&  0.145  &&  0.186      && 0.691       && 0.790        \\
\multicolumn{23}{c}{} \\ [-3mm]
10\% && 6.28 && 2.10   &&  1\% && 0.994 && 0.999 && 0.999  &&  0.739        &&  0.425     &&   0.860      &&  0.950           \\
     &&&& &&  5\%          &&  0.974  && 0.996 && 0.991 &&  0.223   && 0.134    && 0.558        &&   0.804           \\
     &&&& &&  10\%          &&  0.952  && 0.984 && 0.950 &&  0.087   &&  0.050      &&  0.351       &&    0.663        \\
\multicolumn{23}{c}{} \\ [-3mm]
15\%  && 5.19 && 1.80 &&  1\%  && 0.994&& 1.000 && 0.998   &&  0.672        &&   0.283   &&   0.527      &&  0.787           \\
     &&&& &&  5\%           &&  0.974 && 0.997 && 0.992 &&  0.200   && 0.052  && 0.206        &&   0.498           \\
    &&&& &&  10\%           && 0.947 && 0.980 &&  0.945 &&   0.051    &&  0.011      &&  0.090       &&   0.332        \\
\hline
  \end{tabular}\par}
\label{table:sum_FF} 
\end{table}


When $p_0 = 0$ (Panel A of Table \ref{table:sum_FF}), we see that most metrics considered meet the pre-specified significance levels (with the exception of the $90^{th}$ percentile at the 1\% level). This means that when the null hypothesis is true (i.e., no fund is outperforming), the chance of falsely claiming the existence of outperforming funds with a given test statistic is below the pre-specified significance level. This result confirms that the Fama and French (2010) approach performs well in terms of the Type I error rate for the mutual fund data. Notice that this is not a trivial finding in that bootstrap-based methods are not guaranteed to meet the pre-specified significance level (Horowitz (2001)).

When $p_0>0$ (Panel B of Table \ref{table:sum_FF}), we allow some funds to have the ability to generate positive alphas, so a powerful test statistic should be able to detect these outperforming funds and reject the null hypothesis with a high probability. However, this is not the case for the Fama and French (2010) approach. The Type II error rates (i.e., the probability of failing to reject the null hypothesis) are very high. For example, when $p_0 = 2\%$ of funds are truly outperforming, the chance of the best-performing metric (i.e., the $99^{th}$ percentile in this case, since it has the lowest Type II error rate among all seven test statistics) committing a Type II error is 86.9\% under the 5\% significance level. Indeed, even when $p_0 = 5\%$ of funds are outperforming, the lowest Type II error rate across the different test statistics is still 33.6\%.

Our finding of low test power for the Fama and French (2010) method is economically significant. For example, when $p_0= 2\%$, the average alpha and the average $t$-statistic of alpha for outperforming funds is 10.66\% (per annum) and 3.30, respectively. In other words, even when 2\% of funds are truly outperforming and are endowed with on average an annualized alpha of 10.66\%, there is still a 86.9\% chance (at the 5\% significance level) of the Fama and French (2010) approach falsely declaring a zero alpha for all funds.

Note that while our framework proposes a data-driven approach to inject alphas into funds that are assumed to be truly outperforming, an alternative approach is to inject hypothetical alphas directly. For example, Fama and French (2010) assume that true alphas follow a normal distribution and randomly assign alphas from this distribution to funds. We believe that our approach is potentially advantageous in that we do not need to take a stand on the cross-sectional alpha distribution among outperforming funds. For instance, we do not observe extremely large $t$-statistics for alphas (e.g., a $t$-statistic of 10), as large alphas are usually matched with funds with a high level of risk (which is consistent with a risk and return trade-off). However, the random assignment of normally distributed alphas in Fama and French (2010) may assign large alphas to funds with small risk, implying abnormally high $t$-statistics for certain funds that may never be observed in the data. In contrast, our approach injects the alpha for each outperforming fund based on its in-sample alpha estimate. We therefore preserve the data implied cross-sectional alpha distribution for outperforming funds.

Our results also highlight a nonmonotonic pattern in the Type II error rate as a function of the percentile for the test statistic. When $p_0$ is relatively large (e.g., $\geq2\%$), the $98^{th}$ percentile or the $99^{th}$ percentile seem to perform the best in terms of the Type II error rate. In contrast, test statistics with either the highest percentile (e.g., the maximum and the $99.9^{th}$ percentile) or the lowest percentile (i.e., the $90^{th}$ percentile) perform significantly worse.\footnote{
The question of which percentile statistics to use deserves some discussion. In our context, we show that all test statistics have low power so this question does not affect the interpretation of our results. In general, more extreme statistics are likely more powerful. For example, if the true $p_0$ equals 5\%, then percentiles greater than 95\% should be more likely to be rejected than the other percentiles. More extreme statistics are perhaps less interesting, however, from an economic perspective. For instance, if the max statistic rejects but not the $99.9$ th percentile, this would seem to suggest that at most 0.1\% of funds are outperforming, which is not too different from the overall null hypothesis that no fund is outperforming. Taken together, one should balance statistical power with economic significance when choosing test statistics. } Our explanation for this result is that while test statistics with a higher percentile are generally more sensitive to the existence of a small fraction of outperforming funds and therefore should be more powerful, test statistics with the highest percentiles may be sensitive to outliers in $t$-statistics generated by both skilled and unskilled funds.\footnote{For example, suppose we use the maximum $t$-statistic as the test statistic. In the bootstrapped simulations in Fama and French (2010), a fund with a small sample of returns may have an even smaller sample size after bootstrapping, potentially resulting in extreme $t$-statistics for the bootstrapped samples. While this happens for both the (assumed) skilled and unskilled funds, given the larger fraction of unskilled funds, they are more likely to drive the max statistic, making it powerless in detecting outperforming funds. We examine this issue in depth in Section II.C.3, where we dissect the Fama and French (2010) approach. }

How important is our result in the context of performance evaluation for mutual funds? The question of whether outperforming mutual fund managers exist is economically important. Despite its importance, however, there has been considerable debate about the answer to this question. KTWW independently bootstrap each fund's return and find evidence for outperformance for a significant fraction of funds. Fama and French (2010) modify KTWW by taking cross-sectional dependence into account and find no evidence of outperformance. Aside from the different samples in the two papers, we offer a new perspective that helps reconcile these conflicting findings.\footnote{Note that a fund is required to have at least 30 observations in KTWW's analysis. As such, the undersampling issue that we identify with the Fama-French (2010) approach as explained in Section II.C.3 affects KTWW to a much lesser extent.
}

The Fama and French (2010) approach lacks power to detect outperforming funds. Even when 2\% to 5\% of funds are truly outperforming and, moreover, these funds are endowed with the large (and positive) returns that the top 2\% to 5\% of funds observe in the actual sample as in our simulation design, the Fama and French (2010) approach still falsely declares a zero alpha across all funds with a high probability.

Note that the Fama and French (2010) approach, by taking cross-sectional information into account, achieves the pre-specified significance level when the null hypothesis is that all funds generate zero alpha. As such, theoretically, it should make fewer mistakes (i.e., Type I errors) than KTWW when the Fama and French (2010) null hypothesis is true. However, the price we have to pay for making fewer Type I errors is a higher Type II error rate. Our results show that the Type II error rate of the Fama and French (2010) approach is so high that it may mask the existence of a substantial fraction of true outperformers.

In Section II of the Internet Appendix, we follow Fama and French (2010) and perform our analysis on alternative groups of funds classified by fund size. In particular, we examine funds with initial AUM exceeding \$250 million and \$1 billion. The results are consistent with our main results in Table \ref{table:sum_FF}.



Overall, given the emphasis on ``stars" in the investment industry,\footnote{See, for example, Nanda, Wang, and Zheng (2004), Del Guercio and Tkac (2008), and Sastry (2013).} we show that,  in general, it is difficult to identify stars. While a stringent statistical threshold has to be used to control for luck by a large number of unskilled managers, such a threshold makes it hard for good managers, even those with stellar performance, to stand out from the crowd.

\subsubsection{Dissecting the Fama and French Approach}
One key aspect of the Fama and French (2010) bootstrapping approach is that if a fund does not exist for the full sample period, the number of observations for the bootstrapped sample may differ from the number of observations for the actual sample. Harvey and Liu (2020) find that undersampled funds (i.e., funds with fewer observations in the bootstrapped sample than in the actual sample) with a small number of time-series observations in the bootstrapped sample tend to generate $t$-statistics that are more extreme than what the actual sample implies. This distorts the cross-sectional distribution of $t$-statistics for the bootstrapped samples, making it difficult for the Fama and French (2010) approach to reject the null. For details of this analysis, see Harvey and Liu (2020).

\subsubsection{Are There Outperforming Funds? }
\indent Guided by our previous analysis of test power, we now explore ways to improve the performance of the Fama and French (2010) approach and revisit the question of whether outperforming funds exist.

Since the undersampling of funds with a relatively short sample period contributes to the low test power of the Fama and French (2010) approach, the methods we explore in this section all boil down to the question of which funds to drop to alleviate the issue of undersampling. In particular, we explore two ways to drop funds with a small sample size. One way is to focus on the full sample (similar to Fama and French (2010)) but drop funds with a sample size below a threshold $T$ (e.g., $T = 36$). The other way is to focus on different subsamples and require that funds have complete data over these subsamples.

Notice that while dropping funds with a short sample improves the performance of the Fama and French (2010) approach, it raises a concern about sample selection. To the extent that the average performance across funds with a certain sample size is correlated with sample size, our results in this section may provide a biased evaluation of the existence of outperforming funds. For example, while survivorship may bias our results towards finding significant outperforming funds (Brown et al. (1992), Carhart et al. (2002), and Elton, Gruber, and Blake (1996)), reverse-survivorship bias (Linnainmaa (2013)) or decreasing returns to scale (Chen et al. (2004), Berk and Green (2004), Harvey and Liu (2018)) may bias our results in the opposite direction. Because of this concern, we limit the interpretation of our results to the particular sample of mutual funds we analyze, acknowledging that we are not trying to answer exactly the same question as in Fama and French (2010) (i.e., does any fund in their mutual fund sample outperform) since we examine different samples of funds.

For the full-sample analysis similar to Fama and French (2010), we explore two specifications for the threshold $T$: $T = 36$ and $T = 60$. For the subsample analysis, we focus on five-year subsamples and require complete data for each subsample.\footnote{In particular, we split our data into five five-year subsamples (i.e., 1984 to 1988, 1989 to 1993, 1994 to 1998, 1999 to 2003, and 2004 to 2008) and an eight-year subsample (i.e., 2009 to 2016) for the last part of our data.}

%
%
%
%
%
%

We first apply our framework to evaluate test power for the new specifications. Tables IA.VIII and IA.IX in Internet Appendix report the results for the full-sample analysis. Table IA.X reports the average test power across all subsamples, whereas more detailed results on test power for each subsample are presented in Tables IA.XI to IA.XVI. In general, the performance of the Fama and French (2010) approach in terms of test power improves compared to Table \ref{table:sum_FF}, both for the full-sample and the subsample analysis.

We now explore the use of the modified Fama and French (2010) approach to test for the existence of outperforming funds. Table \ref{table:fullsample_FF_results} presents the full-sample results and Table \ref{table:subsample_FF_results} the sub-sample tests.

Focusing on the 1984 to 2006 sample in Table \ref{table:fullsample_FF_results}, the $p$-values for the original Fama-French (2010) approach across test statistics are uniformly higher than those for the adjusted methods. This finding highlights the lack of power of the original Fama-French (2010) approach caused by its missing data bootstrap. Based on the adjusted Fama-French (2010) methods, we find evidence of outperforming funds for the max statistic (at the 1\% level) and the $99.9$th percentile (at the 10\% level), for both $T\geq 36$ and $T \geq 60$.

It is worth emphasizing that the statistical significance of the $\alpha^{th}$ percentile (as the test statistic) may not be solely related to the performance of funds in the top (100-$\alpha$)\%. As a result, it is incorrect to use the significance of a given test statistic to infer the fraction of outperforming funds. Our previous application to the $18,000+$ anomalies makes a similar point. For example, the significance of the max statistic does not imply that only one fund is outperforming. Based on Table IA.IX (i.e., the cutoff $T$ is set to 60), the fact that both the maximum and the $99.9^{th}$ percentile are significant at the 10\% level (as in Table \ref{table:fullsample_FF_results}) while the other test statistics are not may indicate that $p_0 = 0.5\%$. This is because, at $p_0 = 0.5\%$, our simulation results show that the probabilities for the maximum and the $99.9^{th}$ percentile to correctly reject the null hypothesis are 95.6\% ($= 1-0.044$) and 98.2\% ($=1-0.018$), respectively, which are much higher than the probabilities implied by other test statistics (with the highest one being 50.8\% $= 1-0.492$ for the $99.5^{th}$ percentile).

While our full-sample results essentially assume the constancy of alpha for funds, there may be low-frequency fluctuations in fund alphas that are caused by time-varying market conditions or decreasing returns to scale.\footnote{See, for example, Chen et al. (2004), Berk and Green (2004), Avramov et al. (2011), and Harvey and Liu (2018).} We therefore turn to our subsample results in Table \ref{table:subsample_FF_results}. Using the adjusted Fama-French (2010) approach (requiring longer fund histories) and consistent with existing papers that document time-variation in the average performance across funds, we find strong evidence for outperforming funds in the early part of the sample (in particular, the 1984 to 1993 subsample). But there is still some evidence for outperforming funds in the later part of the sample (i.e., the 1999 to 2016 subsample).\footnote{Note that by conditioning on having 60 monthly observations (i.e., no missing observations) for each subsample, we may omit funds that survive only for part of the sample (e.g., 36 months). This could induce a surviorship bias (Elton, Gruber, and Blake (1996)) or a reverse survivorship bias (Linnainmaa (2013)) which would impact our inference.} In contrast, under the original Fama-French (2010) approach, there is essentially no evidence of outperforming funds in any subsample.

Our analysis so far highlights the difference between the Fama and French (2010) approach and the other multiple-testing methods studied in Section II.A, which try to identify the fraction of outperforming funds. While the fraction of outperforming funds (as modeled by $p_0$ in our framework) influences the test power of the Fama and French (2010) approach, in general one cannot infer this fraction from the Fama and French (2010) test statistics. Romano and Wolf (2005) have a similar discussion of this difference. For other methods (including alternative multiple-testing approaches) that can be used to estimate the fraction of outperforming funds, see Barras, Scaillet, and Wermers (2010), Ferson and Chen (2017), Harvey and Liu (2018), and Andrikogiannopoulou and Papakonstantinou (2016). Although these methods can potentially provide a more detailed description of the cross-sectional distribution of alphas, they are not necessarily as powerful as the Fama-French (2010) approach (especially the adjusted Fama-French (2010) approach) in detecting outperforming funds when controlling for cross-sectional dependence in tests. Our focus is on the performance of the Fama and French (2010) approach itself. We leave examination of these alternative methods' test power using our framework to future research.

Overall, the results in this section demonstrate how our framework can be used to evaluate the Fama and French (2010) approach, which is a joint test that is different from the multiple-testing methods that we study in Section II.A. We show that this method has low test power when applied to the actual mutual fund data. Further analysis reveals that the undersampling of funds with a small number of observations produces unrealistically high $t$-statistics and greatly decreases the power to identify outperformers. Accordingly, we modify the Fama and French (2010) approach by dropping funds with a short return history over either the full sample or subsamples. We then revisit the problem in Fama and French (2010) and find some evidence for the existence of outperforming funds. That said, consistent with the long literature in mutual fund evaluation, we find only modest evidence of fund outperformance.

\begin{table}[h!]
\centering 
  \footnotesize\addtolength{\tabcolsep}{-4pt}
  \caption{\textbf{Tests for the Existence of Outperforming Funds: Full Sample}  }
\captionsetup{width=16.0cm}
\caption*{\textmd{\small{This table presents results of tests for the existence of outperforming funds over the full sample.} For a given sample period, we drop funds with the number of observations below a certain threshold and use the Fama and French (2010) approach to test the joint hypothesis of a zero alpha across all funds. For test statistics, we consider six percentiles as well as the maximum. All funds with initial AUM exceeding \$5 million are included. We consider both the Fama and French (2010) sample period (i.e., 1984 to 2006) and the full sample period (i.e., 1984 to 2016). We present $p$-values in parentheses. $^{***}$, $^{**}$, and $^*$ denote statistical significance at the 1\%, 5\%, and 10\% level, respectively.}}
\hspace*{-0.7cm}
\footnotesize{
\begin{tabular}{ccccccccccccccccccccc} 
\hline\hline
      &&      &&             &&          \multicolumn{13}{c}{Test Statistics (for various percentiles)} \\
      \cline{7-21}
Sample Period && \# of Funds && && Max && 99.9\% && 99.5\% && 99\% && 98\% && 95\% && 90\%  \\
\hline
\hline
\multicolumn{21}{c}{} \\ [-2mm]
\multicolumn{21}{c}{Panel A: Benchmark (Fama and French, 2010): $T \geq 8$} \\
\hline
\multicolumn{21}{c}{} \\ [-2mm]
1984-2006   && 3030         &&$t$-stat  && 6.816 && 3.718 && 2.664  && 2.387 && 1.968 && 1.542 && 1.087 \\
(Fama and French) &&             &&$p$-value && \footnotesize{(0.951)} && \footnotesize{(0.939)} &&  \footnotesize{(0.847)}  &&   \footnotesize{(0.737)}      &&   \footnotesize{(0.809)}   &&  \footnotesize{(0.754)}        &&   \footnotesize{(0.841)} \\
          \hline
\multicolumn{21}{c}{} \\ [-2mm]
\multicolumn{21}{c}{Panel B: $T \geq 36$} \\
\hline
\multicolumn{21}{c}{} \\ [-2mm]
1984-2006   && 2,668         &&$t$-stat  && $6.816^{***}$ && $3.759^{*}$ && 2.660  && 2.370 && 1.971 && 1.573 && 1.103 \\
(Fama and French) &&             &&$p$-value && \footnotesize{(0.003)} && \footnotesize{(0.097)} &&  \footnotesize{(0.489)}  &&   \footnotesize{(0.509)}      &&   \footnotesize{(0.643)}   &&  \footnotesize{(0.617)}        &&   \footnotesize{(0.760)} \\
          \multicolumn{21}{c}{} \\ [-2mm]
1984-2016   && 3,868         &&$t$-stat  && $6.959^{***}$ && 3.487 && 2.815  && 2.508 && 2.010 && 1.466 && 1.038 \\
(Full sample) &&             &&$p$-value && \footnotesize{(0.004)} && \footnotesize{(0.289)} &&  \footnotesize{(0.309)}  &&   \footnotesize{(0.337)}      &&   \footnotesize{(0.629)}   &&  \footnotesize{(0.809)}        &&   \footnotesize{(0.884)} \\
          \hline
\multicolumn{21}{c}{} \\ [-2mm]
\multicolumn{21}{c}{Panel C: $T \geq 60$} \\
\hline
\multicolumn{21}{c}{} \\ [-2mm]
1984-2006   && 2,387         &&$t$-stat  && $6.816^{***}$ && $3.824^{*}$ && 2.699  && 2.411 && 2.006 && 1.621 && 1.134 \\
(Fama and French) &&             &&$p$-value && \footnotesize{(0.001)} && \footnotesize{(0.088)} &&  \footnotesize{(0.407)}  &&   \footnotesize{(0.428)}      &&   \footnotesize{(0.577)}   &&  \footnotesize{(0.537)}        &&   \footnotesize{(0.702)} \\
          \multicolumn{21}{c}{} \\ [-2mm]
1984-2016   && 3,393         &&$t$-stat  && $6.959^{***}$ && 3.541 &&  2.897 && 2.546 && 2.089 && 1.529 && 1.075 \\
(Full sample) &&             &&$p$-value && \footnotesize{(0.000)} && \footnotesize{(0.182)} &&  \footnotesize{(0.198)}  &&   \footnotesize{(0.265)}      &&   \footnotesize{(0.472)}   &&  \footnotesize{(0.690)}        &&   \footnotesize{(0.808)} \\
\hline
  \end{tabular}\par}
\label{table:fullsample_FF_results} 
\end{table}

\clearpage

\begin{table}[h!]
\centering 
  \footnotesize\addtolength{\tabcolsep}{-4pt}
  \caption{\textbf{Tests for the Existence of Outperforming Funds: Sub-samples}  }
\captionsetup{width=16.0cm}
\caption*{\textmd{\small{This table presents results of tests for the existence of outperforming funds over subsamples.} For a given sample period, we only keep funds with complete return data over the period and use the Fama and French (2010) approach to test the joint hypothesis of a zero alpha across all funds. For test statistics, we consider six percentiles as well as the maximum. All funds with initial AUM exceeding \$5 million are included. We present $p$-values in parentheses. $^{***}$, $^{**}$, and $^*$ denote statistical significance at the 1\%, 5\%, and 10\% level, respectively.} }
\footnotesize{
\begin{tabular}{ccccccccccccccccccccccc} 
\hline\hline
      &&  &&    &&             &&          \multicolumn{13}{c}{Test Statistics (for various percentiles) } \\
      \cline{9-23}
Subsample && && \# of Funds && && Max && 99.9\% && 99.5\% && 99\% && 98\% && 95\% && 90\%  \\
\hline
1984-88  && $T \geq 8$ && 455         &&$t$-stat  && $4.061$ && $4.061$ && $3.709$  &&   $3.389$       &&  $3.030$   && $2.376$  &&  $1.916$ \\
          && &&              &&$p$-value && \footnotesize{(0.764)} && \footnotesize{(0.764)} &&  \footnotesize{(0.537)}  &&   \footnotesize{(0.365)}      &&   \footnotesize{(0.209)}   &&  \footnotesize{(0.129)}        &&   \footnotesize{(0.111)} \\
\multicolumn{23}{c}{} \\ [-2mm]
         && $T=60$ && 238         &&$t$-stat  && $4.061^{**}$ && $4.061^{**}$ && $4.001^{**}$  &&   $3.464^{**}$       &&  $2.881^{*}$   && $2.375^{*}$  &&  $1.896^{*}$ \\
         && &&             &&$p$-value && \footnotesize{(0.048)} && \footnotesize{(0.048)} &&  \footnotesize{(0.022)}  &&   \footnotesize{(0.032)}      &&   \footnotesize{(0.060)}   &&  \footnotesize{(0.054)}        &&   \footnotesize{(0.058)} \\
\multicolumn{23}{c}{} \\ [-2mm]
1989-93  && $T \geq 8$ && 1,109   &&$t$-stat  && $3.887$ && $3.755$ && $3.183$  &&   $2.820$       &&  $2.459$   && $1.924$  &&  $1.489$ \\
&&    &&   &&$p$-value && \footnotesize{(0.979)} && \footnotesize{(0.973)} &&  \footnotesize{(0.835)}  &&   \footnotesize{(0.755)}      &&   \footnotesize{(0.661)}   &&  \footnotesize{(0.545)}        &&   \footnotesize{(0.456)} \\
\multicolumn{23}{c}{} \\ [-2mm]
   && $T=60$ && 352        &&$t$-stat  && $3.887^{*}$ && $3.887^{*}$ && $3.152^{*}$  &&   $3.011^{*}$       &&  $2.680^{*}$   && $2.179^{*}$  &&  $1.704$ \\
&&       &&      &&$p$-value && \footnotesize{(0.083)} && \footnotesize{(0.083)} &&  \footnotesize{(0.089)}  &&   \footnotesize{(0.087)}      &&   \footnotesize{(0.084)}   &&  \footnotesize{(0.093)}        &&   \footnotesize{(0.116)} \\
\multicolumn{23}{c}{} \\ [-2mm]
1994-98  && $T\geq8$ && 1,857        &&$t$-stat  && $3.482$ && $3.339$ && $2.675$  &&   $2.127$       &&  $1.877$   && $1.437$  &&  $0.919$ \\
&&     &&        &&$p$-value && \footnotesize{(0.992)} && \footnotesize{(0.976)} &&  \footnotesize{(0.906)}  &&   \footnotesize{(0.937)}      &&   \footnotesize{(0.905)}   &&  \footnotesize{(0.893)}        &&   \footnotesize{(0.972)} \\
\multicolumn{23}{c}{} \\ [-2mm]
 && $T=60$ && 848        &&$t$-stat  && $3.237$ && $3.057$ && $2.104$  &&   $1.913$       &&  $1.596$   && $1.182$  &&  $0.692$ \\
&&      &&       &&$p$-value && \footnotesize{(0.497)} && \footnotesize{(0.549)} &&  \footnotesize{(0.859)}  &&   \footnotesize{(0.844)}      &&   \footnotesize{(0.899)}   &&  \footnotesize{(0.940)}        &&   \footnotesize{(0.995)} \\
\multicolumn{23}{c}{} \\ [-2mm]
1999-03  && $T\geq8$   && 2,822         &&$t$-stat  && $5.791$ && $3.797$ && $3.043$  &&   $2.726$       &&  $2.256$   && $1.729$  &&  $1.365$ \\
&&     &&        &&$p$-value && \footnotesize{(0.924)} && \footnotesize{(0.899)} &&  \footnotesize{(0.633)}  &&   \footnotesize{(0.480)}      &&   \footnotesize{(0.494)}   &&  \footnotesize{(0.488)}        &&   \footnotesize{(0.434)} \\
\multicolumn{23}{c}{} \\ [-2mm]
      && $T = 60$   && 1,511         &&$t$-stat  && $3.533$ && $3.508$ && $3.048^{*}$  &&   $2.756$       &&  $2.472$   && $1.874$  &&  $1.508$ \\
&&     &&        &&$p$-value && \footnotesize{(0.346)} && \footnotesize{(0.198)} &&  \footnotesize{(0.087)}  &&   \footnotesize{(0.146)}      &&   \footnotesize{(0.145)}   &&  \footnotesize{(0.238)}        &&   \footnotesize{(0.242)} \\
\multicolumn{23}{c}{} \\ [-2mm]
2004-08  && $T\geq8$ && 3,084         &&$t$-stat  && $4.561$ && $3.667$ && $2.969$  &&   $2.591$       &&  $2.218$   && $1.727$  &&  $1.237$ \\
&&     &&        &&$p$-value && \footnotesize{(0.981)} && \footnotesize{(0.954)} &&  \footnotesize{(0.815)}  &&   \footnotesize{(0.733)}      &&   \footnotesize{(0.660)}   &&  \footnotesize{(0.580)}        &&   \footnotesize{(0.641)} \\
\multicolumn{23}{c}{} \\ [-2mm]
        && $T=60$ && 1,722         &&$t$-stat  && $4.295^{*}$ && $3.594$ && $2.964$  &&   $2.685$       &&  $2.262$   && $1.770$  &&  $1.303$ \\
&&    &&         &&$p$-value && \footnotesize{(0.095)} && \footnotesize{(0.228)} &&  \footnotesize{(0.231)}  &&   \footnotesize{(0.235)}      &&   \footnotesize{(0.316)}   &&  \footnotesize{(0.371)}        &&   \footnotesize{(0.464)} \\
\multicolumn{23}{c}{} \\ [-2mm]
2009-16  && $T\geq8$ && 2,608        &&$t$-stat  && $4.192$ && 3.348 && 2.338  &&   $1.981$       &&  $1.782$   && $1.271$  &&  $0.807$ \\
&&       &&      &&$p$-value && \footnotesize{(0.928)} && \footnotesize{(0.875)} &&  \footnotesize{(0.915)}  &&   \footnotesize{(0.934)}      &&   \footnotesize{(0.876)}   &&  \footnotesize{(0.930)}        &&   \footnotesize{(0.978)} \\
\multicolumn{23}{c}{} \\ [-2mm]
   && $T\geq60$   && 1,642       &&$t$-stat  && $4.192^{*}$ && 3.341 && 2.363  &&   $1.987$       &&  $1.813$   && $1.385$  &&  $0.883$ \\
&&     &&        &&$p$-value && \footnotesize{(0.074)} && \footnotesize{(0.240)} &&  \footnotesize{(0.614)}  &&   \footnotesize{(0.764)}      &&   \footnotesize{(0.685)}   &&  \footnotesize{(0.741)}        &&   \footnotesize{(0.903)} \\
\hline
  \end{tabular}\par}
\label{table:subsample_FF_results} 
\end{table}

\section{Conclusion}
\indent Two types of mistakes (i.e., false discoveries and missed discoveries) vex empirical research in financial economics. Both are exacerbated in the context of multiple tests. We propose a data-driven approach that allows us to estimate the frequencies of both errors for multiple tests. Our approach can also be used to flexibly estimate functions of the two frequencies, such as a weighted average of the false discovery frequency and the missed discovery frequency, with the weight determined by the cost of the type of mistake.

While current research on multiple testing focuses on controlling the Type I error rate, we show that it is also important to consider the Type II error rate. For the selection of investment strategies, a weighted average of the Type I error rate and the Type II error rate is likely more consistent with the investor's objective function. For the selection of mutual fund managers, current methods, which ignore the Type II error rate, may lead us to miss outperforming managers. Instead of relying on existing multiple-testing adjustments, our approach allows for the provision of user-specific significance thresholds for a particular set of data. Alternatively, if the researcher wants to use a traditional multiple-testing adjustment, our method is able to determine which one is best suited to the particular application at hand.

With the advent of big data and advances in computing technologies, it becomes increasingly important to correct the biases associated with multiple tests and data mining. Taking advantage of current computing technologies, we develop a simulation-based framework to make such corrections. We expect our method to be useful for future research in financial economics as well as other fields.

\singlespacing

\clearpage

\setcounter{figure}{0} \renewcommand{\thefigure}{A.\Roman{figure}}
\setcounter{table}{0} \renewcommand{\thetable}{A.\Roman{table}}

\section*{Appendix: Implementing Romano, Shaikh, and Wolf (2008)}
\indent Romano, Shaikh, and Wolf (2008) suggest a bootstrapping method to take the dependence structure in the data into account to derive the statistical cutoff. Similar to the implementation in Romano, Shaikh, and Wolf (2008), who set the number of bootstrapped iterations to $B = 500$, we set $B = 1,000$ for the bootstrap procedure.

Romano, Shaikh, and Wolf (2008) is computationally intensive in that we have to run $B\times O(M^2)$ regression models to derive the $t$-statistic threshold for a given sample, where $M$ is the total number of tests. This makes it difficult to implement the test for the 18,000 anomalies. We therefore randomly sample $N = 100$ times from the 18,113 anomalies, each time drawing 500 anomalies (which is similar to the size of the CAPIQ data). We then average across these random samples to evaluate the performance of Romano, Shaikh, and Wolf (2008).

\begin{table}[H]
\centering 
  \footnotesize\addtolength{\tabcolsep}{-4pt}
  \captionsetup{justification=centering}
  \caption{\textbf{Simulated Error Rates for Romano, Shaikh, and Wolf (2008): CAPIQ and 18,000 Anomalies}  }
\captionsetup{width=15.0cm}
  \captionsetup{justification=justified}
\caption*{\textmd{\small{This table presents simulated Type I and Type II error rates for CAPIQ and the 18,000 anomalies.} For CAPIQ, for each fraction $p_0$ of strategies that are believed to be true, we follow our method in Section II and set $I = 100$ (for each $i$, we bootstrap to obtain the ranking of strategies and set the top $p_0$ as true) and $J = 100$ (conditional on $i$, for each $j$, we bootstrap the time periods) to run10,000 ($=100\times 100$) bootstrapped simulations to calculate the empirical Type I and Type II error rates for Romano, Shaikh, and Wolf (2008). For the 18,113 anomalies, we first randomly sample 500 strategies $N = 100$ times. We then calculate the empirical Type I and Type II error rates for each random sample, and average across these random samples to obtain the average Type I and Type II error rates. * indicates the highest (and oversized) Type I error rate among all methods considered in Table \ref{table:exist_IQ} and Table \ref{table:exist_YanZheng}.} }
\hspace*{-1cm}
\small{
\begin{tabular}{cccccccccccc} 
\hline\hline
      && &&\multicolumn{3}{c}{Type I} &&& \multicolumn{3}{c}{Type II} \\
      \cline{5-7} \cline{10-12}
$p_0$ && $\alpha$ && CAPIQ  && 18,000 anomalies &&& CAPIQ  &&  18,000 anomalies \\
(frac. of true) && (sig. level) && && &&& &&  \\
\multicolumn{12}{c}{} \\ [-3mm]
2\%   &&  1\%     && 0.017  &&  0.021$^*$  &&& 0.002 && 0.008 \\
      &&  5\%     && 0.045  &&  0.084$^*$  &&& 0.001 && 0.006 \\
      &&  10\%    && 0.142$^*$  &&  0.146$^*$  &&& 0.001 && 0.005    \\
      \multicolumn{12}{c}{} \\ [-3mm]
5\%   &&  1\%     && 0.008  && 0.020$^*$  &&& 0.005 && 0.028 \\
      &&  5\%     && 0.054  && 0.080$^*$  &&& 0.002 &&  0.019  \\
      &&  10\%    && 0.119$^*$  && 0.146$^*$  &&& 0.001 &&  0.016   \\
      \multicolumn{12}{c}{} \\ [-3mm]
10\%   &&  1\%     && 0.008  &&  0.016 &&& 0.019 && 0.070 \\
      &&  5\%     && 0.060  &&  0.074$^*$  &&& 0.007 && 0.050  \\
      &&  10\%    && 0.125$^*$ && 0.129$^*$  &&& 0.004 &&  0.049  \\
      \multicolumn{12}{c}{} \\ [-3mm]
20\%   &&  1\%     && 0.006 && 0.021$^*$  &&& 0.078 && 0.168  \\
      &&  5\%     && 0.055  &&  0.057 &&& 0.039 && 0.137 \\
      &&  10\%    && 0.114  && 0.120$^*$  &&& 0.026 && 0.111   \\
            \multicolumn{12}{c}{} \\ [-3mm]
30\%   &&  1\%     && 0.004 &&  0.018$^*$ &&& 0.179 &&  0.265 \\
      &&  5\%     && 0.047  &&  0.060 &&& 0.115 &&  0.244 \\
      &&  10\%    && 0.093  && 0.113  &&& 0.091 &&  0.201  \\
\hline
  \end{tabular}\par}
\label{table:RSW} 
\end{table}

\end{document}